\documentclass[12pt,leqno]{article}
\title{Gravitational instability of finite isothermal spheres in general relativity. Analogy with neutron stars }
\author{Pierre-Henri Chavanis$^{1,2}$}
\date{}

\usepackage{amsthm,amsmath,amssymb,a4wide,psfig}
\def\mb#1{\setbox0=\hbox{$#1$}\kern-.025em\copy0\kern-\wd0
\kern-0.05em\copy0\kern-\wd0\kern-.025em\raise.0233em\box0}

\begin{document}
\maketitle
\vspace*{-1cm}
\begin{center}
$^{1}$ Laboratoire de Physique Quantique,
Universit\'e Paul Sabatier,\\
118 route de Narbonne 31062 Toulouse, France.\\

$^{2}$ Institute for Theoretical Physics,
University of California, Santa Barbara, California.\\

\vspace{0.5cm}
\end{center}

\begin{abstract}

We investigate the effects of relativity on the gravitational
instability of finite isothermal gaseous spheres. In the first part of
the paper, we treat the gravitational field within the framework of
Newtonian mechanics but we use a relativistic equation of state in the
condition of hydrostatic equilibrium. In the second part of the paper,
we study the full general relativistic problem for a gas described by
an equation of state $p=q\epsilon$ such that the pressure is
proportional to the energy density (``isothermal'' distribution). For
$q=1/3$, this equation of state describes the core of neutron
stars. The mass-density diagram displays some damped oscillations and
there exists a critical value of mass-energy above which no
equilibrium state is possible. We show analytically that the mass
peaks are associated with new modes of instability. These results are
strikingly similar to those obtained by Antonov
[Vest. Leningr. Gos. Univ. 7, 135 (1962)] and Lynden-Bell \& Wood
(1968) for a classical isothermal gas. Our study completes the analogy
between isothermal spheres and neutron stars investigated by Yabushita
[MNRAS 167, 95 (1974)].

\end{abstract}


\section{Introduction}
\label{sec_introduction}

Isothermal spheres play an important role in astrophysics. They were
initially introduced in the context of stellar structure
(Chandrasekhar 1942) when composite configurations of stars consisting
of an isothermal core and a polytropic envelope were constructed and
studied. On larger scales, they were applied to stellar systems such
as globular clusters and elliptical galaxies (Binney \& Tremaine
1987). The age of globular clusters is such that their isothermal
structure is due to a succession of encounters between stars that lead
to an equipartition of energy, like in an ordinary gas. This
statistical mechanics prediction has been confirmed by direct
observations of globular clusters and is well reproduced by King's
models that incorporate a truncation in the distribution function so
as to account for tidal effects. In the case of elliptical galaxies
(and possibly other collisionless self-gravitating systems like Dark
Matter), the isothermal distribution is a result of a ``violent
relaxation'' by phase mixing (Lynden-Bell 1967). In that case, there
is no segregation by mass contrary to the collisional relaxation. In
general, violent relaxation is incomplete and the distribution function
must be modified at high energies. However, the distribution function
that prevails in the inner regions of elliptical galaxies is
isothermal and this is an important ingredient to understand de
Vaucouleurs' $R^{1/4}$ law (Hjorth \& Madsen 1993). Isothermal
distributions are also appropriate to describe the cold interstellar
medium where the temperature is imposed by the cosmic background
radiation at $T\sim 3K$ in the outer parts of galaxies, devoid of any
star and heating sources (Pfenniger
\& Combes 1994, de Vega {\it et al.} 1996). They can also be of
interest in cosmology to understand the fractal structure of the
universe (Saslaw
\& Hamilton 1984, de Vega {\it et al.} 1998).

The stability of isothermal gaseous spheres in Newtonian gravity has
been investigated by Antonov (1962), Lynden-Bell \& Wood (1968) and
Katz (1978) by using thermodynamical arguments and topological
properties of the equilibrium phase diagram by analogy with
Poincar\'e's theory of linear series of equilibrium. They showed in
particular that, if the energy or the temperature are below a certain
threshold, no hydrostatic equilibrium can exist. In that case, the
system is expected to undergo a phase transition and collapse. This is
called ``gravothermal catastrophe'' when the energy is kept fixed
(microcanonical ensemble) and ``isothermal collapse'' when the system
evolves at a fixed temperature (canonical ensemble).  The
thermodynamics of self-gravitating systems was reconsidered by
Padmanabhan (1989) who calculated explicitly the second order
variations of entropy and reduced the problem of stability to an
eigenvalue equation. His method was applied in the canonical ensemble
by Chavanis (2001) who showed in addition the equivalence between
thermodynamical stability and dynamical stability based on the
Navier-Stokes equations (Jeans problem). Similar studies have been
performed by Semelin {\it et al.} (1999,2000) and de Vega \& Sanchez
(2001) by using field theoretical methods. As shown by Padmanabhan
(1989) and Chavanis (2001), the problem of stability can be studied
without approximation almost analytically (or with graphical
constructions) by making use explicitly of the Milne variables
introduced long ago in the context of stellar structure (Chandrasekhar
1942).

We show in the present paper that these methods naturally extend to
the context of general relativity. Isothermal gaseous spheres (in the
sense given below) have been only poorly studied in general relativity
despite their similarity with classical isothermal spheres. The most
extended study that we have found is the contribution of Chandrasekhar
at the conference given in the honour of J.L. Synge in 1972. His work
was completed by Yabushita (1973,1974) who considered the
stability of a relativistic isothermal gas  surrounded by an external
envelope imposing a constant pressure (an extension of the classical
Bonnor (1956) problem). In order to make the link with the works of
Antonov (1962) and Lynden-Bell \& Wood (1968) in Newtonian gravity, we shall
consider the situation in which the volume is fixed instead of the
pressure. This study can have direct applications to the
stability of neutron stars since the equation of state that prevails
in the central region of these highly relativistic objects is very close
to the ``isothermal'' one.

In section \ref{sec_antonov}, we consider the stability of isothermal
gas spheres described by a relativistic equation of state but still
using the framework of Newtonian gravity. This is a first attempt to
introduce relativistic effects in the classical Antonov problem. In
fact, as is well-known, special relativity does not modify the
equation of state for a perfect gas (Chandrasekhar 1942). Only does it
alter the onset of instability. We find that the critical energy below
which no hydrostatic equilibrium is possible depends on a relativistic
parameter $\mu$ defined as the ratio between the size of the domain
$R$ and a ``classical'' Schwarzschild radius
$R_{S}^{class}=2GM/c^{2}$, where $M$ is the total mass of the
system. In the classical limit $R\gg R_{S}^{class}$, we recover the
Antonov result $E_{c}=-0.335GM^{2}/R$ but when relativistic
corrections are included we find that the critical energy is
increased, i.e., instability occurs {\it sooner} than in the classical
case. The density perturbation profile that triggers the instability
is calculated explicitly. For $\mu\rightarrow +\infty$, it presents a
``core-halo'' structure but relativistic effects tend to reduce the
extent of the halo.  When the system is maintained at a fixed
temperature instead of a fixed energy (canonical description), the
results are unchanged with respect to the Newtonian case.

Of course, when the Schwarzschild radius becomes comparable to the
size of the system, general relativistic effects must be taken into
account. This problem is treated in detail in section
\ref{sec_iso}. We consider a simple equation of state $p=q\epsilon$
(where $q$ is a constant) which generalizes the equation of state for
isothermal spheres in the Newtonian context (the classical limit is
recovered for $q\rightarrow 0$). Since the equations governing
equilibrium have the same structure and the same properties as in the
classical case, we shall say that the system is ``isothermal''
(following the terminology of Chandrasekhar 1972), although this
equation of state does not correspond to thermal equilibrium in a
strict sense (see section
\ref{sec_state}). True statistical equilibria in general relativity have
been investigated by Katz {\it et al.} (1975) and they are
characterized by a non uniform temperature (because of the
gravitational redshift). However, systems described by the equation of
state $p=q\epsilon$ are numerous in nature and they include for
example the important case of neutron cores which are usually modeled
as an assembly of cold degenerate fermions for which $q=1/3$. This
simple isothermal equation of state is the high-density limit of more
general equations of state usually considered for neutron stars
(Oppenheimer
\& Volkoff 1939, Meltzer \& Thorne 1966). Quite remarkably, our simple
``box'' model is able to reproduce the main properties of these objects. This
suggests that the structure of neutron stars is due intrinsically to
their isothermal cores and not to the details of their envelopes (which
are not well-known).  In section \ref{sec_osci}, we show that the
mass-density diagram displays an infinity of mass peaks and that isothermal 
spheres exist only below a
limiting mass corresponding to the first peak. In section
\ref{sec_mr}, we show that the mass-radius diagram has a spiral
behavior similar to the one observed for neutron stars. Using an
equation of pulsation derived by Yabushita (1973), we demonstrate
analytically that the series of equilibrium becomes unstable precisely
at the point of maximum mass (like in the study of Misner \& Zapolsky
1964) and that new modes of instability correspond to secondary mass
peaks. We obtain the same stability criterion from energy
considerations based on the binding energy $E=M-Nmc^{2}$. Said
differently, these results indicate that, for a fixed mass, the system
becomes unstable when its radius is smaller than a multiple of the
Schwarzschild radius, a property consistent with Chandrasekhar's
(1964) general theory. The perturbation profiles of density and
velocity are calculated explicitly and expressed in terms of the Milne
variables. They do not present a ``core-halo'' structure. All these
properties are strikingly similar to those obtained for classical
isothermal spheres in the canonical ensemble (see, e.g., Chavanis
2001). This completes the analogy between isothermal spheres and
neutron stars investigated by Yabushita (1974).

\section{Antonov instability for a relativistic gas sphere}
\label{sec_antonov}  

\subsection{The relativistic gas}
\label{sec_relativistic}

We consider a system of $N$ particles, each of mass $m$,  in gravitational interaction. We allow the speed of the particles to be close to the velocity of light so that special relativity must be taken into account. However, in this first approach, we shall treat the gravitational field within the framework of Newtonian mechanics. This procedure is permissible if the typical size of the system is much larger than the Schwarzschild radius (see, e.g, Chandrasekhar 1942). Let $f({\bf r},{\bf p},t)$ denote the distribution function of the system, i.e. $f({\bf r},{\bf p},t)d^{3}{\bf r}d^{3}{\bf p}$ gives the average mass of particles whose positions and momenta are in the cell $({\bf r},{\bf p}; {\bf r}+d^{3}{\bf r},{\bf p}+d^{3}{\bf p})$ at time $t$. The integral of $f$ over the momenta determines the spatial density
\begin{equation}
\rho=\int f d^{3}{\bf p},
\label{R1}
\end{equation} 
and the total mass is expressed as
\begin{equation}
M=\int \rho d^{3}{\bf r}.
\label{R2}
\end{equation} 
In a meanfield approximation, the total energy is given by
\begin{equation}
E=\int {f\over m} \epsilon \ d^{3}{\bf r}d^{3}{\bf p}+{1\over 2}\int\rho\Phi d^{3}{\bf r}=K+W,
\label{R3}
\end{equation}   
where $K$ and $W$ are the kinetic and potential energy respectively. According to the theory of special relativity, the energy of a particule reads
\begin{equation}
\epsilon=m c^{2}\biggl\lbrace \biggr (1+{p^{2}\over m^{2}c^{2}}\biggr )^{1/2}-1\biggr \rbrace.
\label{R4}
\end{equation}
This expression does not include the rest mass, so it reduces to the usual kinetic energy ${p^{2}\over 2m}$ in the Newtonian limit. All the effects of gravity are incorporated in the potential energy which contains the gravitational potential $\Phi$ related to the density by the Newton-Poisson equation 
\begin{equation}
\Delta\Phi=4\pi G\rho.
\label{R5}
\end{equation}

We now ask which configuration maximizes the Boltzmann entropy 
\begin{equation}
S=-k\int {f\over m}\ln {f\over m} d^{3}{\bf r}d^{3}{\bf p},
\label{R6}
\end{equation}
subject to the conservation of mass $M$ and energy $E$. To that purpose, we proceed in two steps. In this section, we maximize $S[f]$ at fixed $E$, $M$ and $\rho({\bf r})$. This provides an optimal distribution function expressed in terms of $\rho$. Then, in section \ref{sec_first}, we maximize $S_{max}[\rho]$ at fixed $M$ and $E$. This optimization problem is not trivial and will be further discussed in sections \ref{sec_equi}-\ref{sec_cond}. 

Since the gravitational potential can be deduced from the density by solving the Poisson equation (\ref{R5}), maximizing  $S$ at fixed $E$, $M$ and $\rho({\bf r})$ is equivalent to maximizing $S$ at fixed $\rho({\bf r})$ and $K$.  Writing the variational principle in the form
\begin{equation}
\delta S-k\beta\delta K-\int\lambda({\bf r})\delta\rho d^{3}{\bf r}=0,
\label{R7}
\end{equation}
we obtain the optimal distribution function
\begin{equation}
f({\bf r},{\bf p})=A({\bf r})e^{-\beta \epsilon},
\label{R8}
\end{equation}
which is a {\it global} entropy maximum with the previous constraints. Eq. (\ref{R8}) is the relativistic Maxwell-Boltzmann distribution with an inverse temperature
\begin{equation}
\beta={1\over kT}.
\label{R9}
\end{equation}
The Lagrange multipliers $A({\bf r})$ and $\beta$ must be related to the constraints $\rho({\bf r})$ and $K$. As discussed in detail by Chandrasekhar (1942), these relations can be expressed in terms of the modified Bessel functions
\begin{equation}
K_{n}(z)=\int_{0}^{+\infty}e^{-z\cosh\theta}\cosh(n\theta)d\theta.
\label{R10}
\end{equation}
Using Eqs. (\ref{R1}) (\ref{R8}) (\ref{R4}) and introducing the Juttner transformation ${p\over mc}=\sinh\theta$ in the integral, the density can be written
\begin{equation}
\rho({\bf r})={4\pi m^{3} c^{3}\over x}A({\bf r})K_{2}(x)e^{x}, 
\label{R11}
\end{equation}
where we have introduced the dimensionless parameter
\begin{equation}
x=\beta mc^{2}={mc^{2}\over kT}.
\label{R12}
\end{equation}
This parameter quantifies the importance of relativistic effects. The classical limit corresponds to $x\rightarrow +\infty$ ($kT\ll mc^{2}$) and the ultra-relativistic limit to $x\rightarrow 0$ ($kT\gg mc^{2}$). 
 
The distribution function (\ref{R8}) can now be expressed in terms of the density as
\begin{equation}
f({\bf r},{\bf p})={x \over  4\pi m^{3}c^{3}}  {e^{-x}\over K_{2}(x)}\rho({\bf r}) e^{-\beta\epsilon}. 
\label{R13}
\end{equation}
In the classical limit $(x\rightarrow +\infty)$ we recover the standard formula
\begin{equation}
f({\bf r},{\bf p})={1\over (2\pi mkT)^{3/2}}\rho({\bf r})e^{-{p^{2}\over 2mkT}},
\label{R14}
\end{equation}
and in the  ultra-relativistic limit ($x\rightarrow 0$), we get
\begin{equation}
f({\bf r},{\bf p})={c^{3}\over 8\pi k^{3}T^{3}}\rho({\bf r})e^{-{pc\over kT}}.
\label{R15}
\end{equation}

Similarly, after some elementary transformations, the kinetic energy can be expressed in terms of the normalized inverse temperature $x$ by 
\begin{equation}
K={\cal F}(x)Mc^{2},\qquad {\rm with}\qquad {\cal F}(x)={3K_{3}(x)+K_{1}(x)\over 4K_{2}(x)}-1.
\label{R16}
\end{equation}
Using the recursion formula
\begin{equation}
K_{n-1}(z)-K_{n+1}(z)=-{2n\over z}K_{n}(z),
\label{R17}
\end{equation}
the function ${\cal F}(x)$ can be written in the equivalent form
\begin{equation}
{\cal F}(x)={K_{1}(x)\over K_{2}(x)}+{3\over x}-1.
\label{R18}
\end{equation}
It has the asymptotic behaviors
\begin{equation}
{\cal F}(x)\sim {3\over 2x}\qquad (x\rightarrow +\infty),
\label{R20}
\end{equation}
\begin{equation}
{\cal F}(x)\sim {3\over x}\qquad (x\rightarrow 0).
\label{R21}
\end{equation}
When substituted in Eq. (\ref{R16}), we recover the usual expressions for the kinetic energy $K={3\over 2}NkT$ in the classical limit  and $K=3NkT$ in the ultra-relativistic limit.

We can now express the entropy in terms of the density $\rho$ and the inverse temperature $x$. Substituting the optimal distribution function (\ref{R13}) in Eq. (\ref{R6}), we get, up to an additional constant 
\begin{equation}
S=kN{\cal G}(x)-k\int{\rho\over m}\ln {\rho\over m} d^{3}{\bf r},
\label{R22}
\end{equation}
where
\begin{equation}
{\cal G}(x)=x\lbrack {\cal F}(x)+1\rbrack +\ln K_{2}(x) -\ln x.
\label{R23}
\end{equation}
The function ${\cal G}(x)$ has the asymptotic behaviors
\begin{equation}
{\cal G}(x)\sim -{3\over 2}\ln x \qquad (x\rightarrow +\infty),
\label{R24}
\end{equation}
\begin{equation}
{\cal G}(x)\sim -3\ln x \qquad (x\rightarrow 0).
\label{R25}
\end{equation}
The thermal contribution to the entropy in the classical limit is $S_{th}={3\over 2}kN\ln T$ and in the ultra-relativistic limit $S_{th}={3}kN\ln T$. 

There exists a general relation between the derivatives of ${\cal F}$ and ${\cal G}$ that we shall need in the following. Differentiating Eq. (\ref{R23}) with respect to $x$ and using the identity
\begin{equation}
K'_{n}(x)=-K_{n-1}(x)-{n\over x}K_{n}(x),
\label{R26}
\end{equation}
for $n=2$, we find that
\begin{equation}
{\cal G}'(x)=x{\cal F}'(x).
\label{R27}
\end{equation}

\subsection{First and second order variations of entropy}
\label{sec_first}

In the preceding section, we have expressed the entropy and the kinetic energy in terms of the density $\rho({\bf r})$ and the temperature $T$ (through the variable $x$). We now wish to maximize the entropy $S[\rho]$ at fixed $E$ and $M$. For convenience, we shall introduce a new variable $y={\cal F}(x)$. In terms of this variable, the total energy and the entropy can be written 
\begin{equation}
E=yMc^{2}+{1\over 2}\int\rho\Phi d^{3}{\bf r},
\label{F1}
\end{equation}
\begin{equation}
S=kN{\cal G}(x(y))-k\int{\rho\over m}\ln {\rho\over m} d^{3}{\bf r}.
\label{F2}
\end{equation}
We can now determine the variations of $S$ around a given density profile $\rho({\bf r})$. To second order in the expansion, we get 
\begin{equation}
\delta S=kN{d{\cal G}\over dy}\delta y+kN {d^{2}{\cal G}\over dy^{2}}{(\delta y)^{2}\over 2}-{k\over m}\int \delta\rho \biggl (1+\ln {\rho\over m}\biggr ) d^{3}{\bf r}-{k\over m}\int {(\delta\rho)^{2}\over 2\rho}d^{3}{\bf r}.
\label{F3}
\end{equation}
Using the identity (\ref{R27}), we find that
\begin{equation}
{d{\cal G}\over dy}={d{\cal G}\over dx}{dx\over dy}={{\cal G}'(x)\over {\cal F}'(x)}=x.
\label{F4}
\end{equation}
Differentiating one more time with respect to $y$, we obtain
\begin{equation}
{d^{2}{\cal G}\over dy^{2}}={dx\over dy}={1\over {\cal F}'(x)}.
\label{F5}
\end{equation}
Substituting the above results in Eq. (\ref{F3}), we get
\begin{equation}
\delta S=kN x\delta y+kN{1\over {\cal F}'(x)}  {(\delta y)^{2}\over 2}-{k\over m}\int \biggl (1+\ln {\rho\over m}\biggr )  \delta\rho d^{3}{\bf r}-{k\over m}\int {(\delta\rho)^{2}\over 2\rho}d^{3}{\bf r}.
\label{F6}
\end{equation}
We now need to express the variation $\delta y$ in terms of $\delta\rho$. From the  conservation of energy, we have the exact identity
\begin{equation}
0=\delta E=Mc^{2}\delta y+\int \Phi\delta \rho d^{3}{\bf r}+{1\over 2}\int\delta\rho\delta \Phi d^{3}{\bf r}.
\label{F7}
\end{equation}
Substituting the foregoing expression for $\delta y$ from Eq.  (\ref{F7}) in Eq. (\ref{F6}), we obtain
\begin{eqnarray}
\delta S=-{1\over T}\int  \Phi\delta\rho d^{3}{\bf r}-{1\over 2T}\int\delta\rho\delta\Phi d^{3}{\bf r}+{k\over 2Mmc^{4}}{1\over {\cal F}'(x)}\biggl (\int\Phi\delta\rho d^{3}{\bf r}\biggr )^{2}\nonumber\\
-{k\over m}\int \biggl (1+\ln {\rho\over m}\biggr )\delta\rho d^{3}{\bf r}-{k\over m}\int {(\delta\rho)^{2}\over 2\rho}d^{3}{\bf r}.
\label{F8}
\end{eqnarray}

Introducing a Lagrange multiplier $\alpha$ to satisfy the conservation of mass, the condition that $S$ is an extremum is written (to first order)
\begin{eqnarray}
0=\delta S-\alpha\delta M=-\int\biggl\lbrack  {\Phi\over T} +{k\over m}\biggl (1+\ln {\rho\over m}\biggr )+\alpha \biggr\rbrack \delta\rho d^{3}{\bf r}.
\label{F9}
\end{eqnarray}
This condition must be satisfied for any variations $\delta\rho$. This yields the Boltzmann distribution
\begin{eqnarray}
\rho=Ae^{-{m\Phi\over kT}},
\label{F10}
\end{eqnarray}
like for a classical gas. The condition that the critical point (\ref{F10}) is an entropy {\it maximum} requires that
\begin{eqnarray}
\delta^{2} S=-{1\over 2T}\int\delta\rho\delta\Phi d^{3}{\bf r}+{k\over 2Mmc^{4}}{1\over {\cal F}'(x)}\biggl (\int\Phi\delta\rho d^{3}{\bf r}\biggr )^{2}-{k\over m}\int {(\delta\rho)^{2}\over 2\rho}d^{3}{\bf r}<0
\label{F11}
\end{eqnarray}
for any variation $\delta\rho$ that conserves mass to first order.

\subsection{The Virial theorem for a relativistic gas}
\label{sec_virial}

In this section, we derive the form of the Virial theorem appropriate to an isothermal gas described by a relativistic equation of state. Quite generally, the potential energy of a self-gravitating system can be expressed in the form (Binney \& Tremaine 1987)
\begin{eqnarray}
W=-\int\rho {\bf r}\nabla\Phi d^{3}{\bf r}.
\label{V1}
\end{eqnarray} 
If the system is in hydrostatic equilibrium, then
\begin{eqnarray}
\nabla p=-\rho\nabla\Phi.
\label{V2}
\end{eqnarray}
Substituting this identity in Eq. (\ref{V1}) and integrating by parts, we get
\begin{eqnarray}
W=\int {\bf r}\nabla p d^{3}{\bf r}=\oint p{\bf r}d{\bf S}-3\int p d^{3}{\bf r}.
\label{V3}
\end{eqnarray} 
If the pressure $p_{b}$ on the boundary of the system is uniform, we can write
\begin{eqnarray}
\oint p{\bf r}d{\bf S}=p_{b}\oint {\bf r}d{\bf S}=p_{b}\int \nabla .{\bf r} d^{3}{\bf r}=3p_{b}V,
\label{V4}
\end{eqnarray} 
where $V$ is the total volume of the system. Therefore, for any system in hydrostatic equilibrium, one has
\begin{eqnarray}
W=3p_{b}V-3\int pd^{3}{\bf r},
\label{V5}
\end{eqnarray} 
which can be considered as the general form of the Virial theorem for self-gravitating systems. 

Now, the pressure of an ideal gas  can be expressed as (Chandrasekhar 1942)
\begin{eqnarray}
p={1\over 3}\int {f\over m}p{\partial\epsilon\over\partial p}d^{3}{\bf p}.
\label{V6}
\end{eqnarray}
For a relativistic gas described by the distribution function (\ref{R8}), we get
\begin{eqnarray}
p=-{4\pi\over 3 m\beta} A({\bf r})\int_{0}^{+\infty}{\partial\over\partial p}(e^{-\beta\epsilon})p^{3}dp.
\label{V7}
\end{eqnarray}
Integrating by parts, we obtain
\begin{eqnarray}
p={\rho\over m\beta}={\rho\over m}{k}T.
\label{V8}
\end{eqnarray}
Therefore, the equation of state for a (non quantum) relativistic gas is the same as for a classical gas. Written in the form 
\begin{eqnarray}
W+3NkT=3p_{b}V,
\label{V9}
\end{eqnarray} 
the Virial theorem (\ref{V5}) also has the same form as for an isothermal  classical gas. However, the Virial theorem is usually expressed in terms of the kinetic energy instead of the temperature. Therefore, the appropriate form of the relativistic Virial theorem reads
 \begin{eqnarray}
W+{3Mc^{2}\over {\cal F}^{-1}({K\over Mc^{2}})}=3p_{b}V.
\label{V10}
\end{eqnarray} 
In the classical limit ($x\rightarrow +\infty$), it reduces to the well-known formula
\begin{eqnarray}
W+2K=3p_{b}V,
\label{V11}
\end{eqnarray} 
and in the ultra-relativistic limit ($x\rightarrow 0$), we get
\begin{eqnarray}
W+K=3p_{b}V.
\label{V12}
\end{eqnarray} 
It should be emphasized that Eq. (\ref{V10}) is  only valid for a relativistic gas in {\it thermal} equilibrium.

\subsection{The equilibrium phase diagram}
\label{sec_equi}

Since the equation of state for a relativistic gas is the same as for a classical gas, the equilibrium configurations of such systems correspond to the isothermal gas spheres described extensively in the monograph of Chandrasekhar (1942). Only the {\it onset} of the gravitational instability will be modified by relativistic effects.

For non rotating systems, the equilibrium states are expected to be spherically symmetric. In that case, the Poisson equation (\ref{R5}) together with the Boltzmann distribution (\ref{F10}) yield the second order differential equation
\begin{eqnarray}
{1\over r^{2}}{d\over dr}\biggl (r^{2}{d\Phi\over dr}\biggr )=4\pi GAe^{-\beta m\Phi}.
\label{E1}
\end{eqnarray} 
This equation can also be deduced from the condition of hydrostatic equilibrium (\ref{V2}) when the pressure is related to the density according to the equation of state (\ref{V8}). It is well-known that the density profile of such isothermal configuations  behaves like $\rho\sim r^{-2}$ at large distances so their total mass is infinite. Following Antonov (1962), we shall avoid this infinite mass problem by confining artificially the system within a spherical box of radius $R$. It is only under this simplifying assumption that a rigorous thermodynamics of self-gravitating systems can be carried out (see, e.g., Padmanabhan 1990, Chavanis {\it et al.} 2001). This procedure is justified physically by the realization that a distribution of matter never extends to infinity so $R$ represents an upper cut-off at which other processes intervene to limitate the spatial extent of the system. Of course, different cut-offs are possible but fixing the volume is consistent with the traditional viewpoint of statistical mechanics and it is sufficient to capture the essential physics of the system (see the different comparisons of truncated models performed by Katz (1980)). 

We now wish to determine the equilibrium phase diagram of a relativistic isothermal gas. To that purpose, we introduce the function $\psi=\beta m (\Phi-\Phi_{0})$ where $\Phi_{0}$ is the gravitational potential at $r=0$. Then, the density field (\ref{F10}) can be written
\begin{equation}
\rho=\rho_{0}e^{-\psi},
\label{E2}
\end{equation}
where $\rho_{0}$ is the central density. Introducing the notation $\xi=(4\pi G\beta m \rho_{0})^{1/2}r$, the Boltzmann-Poisson equation (\ref{E1}) reduces to the standard Emden form
\begin{equation}
{1\over\xi^{2}}{d\over d\xi}\biggl (\xi^{2}{d\psi\over d\xi}\biggr )=e^{-\psi}.
\label{E3}
\end{equation}
Eq. (\ref{E3}) has a simple analytic solution, the singular sphere
\begin{equation}
e^{-\psi_{s}}={2\over\xi^{2}},
\label{E4}
\end{equation}
whose central density is infinite. The regular solutions of Eq. (\ref{E3}) must satisfy the boundary conditions
\begin{equation}
\psi(0)=\psi'(0)=0,
\label{E5}
\end{equation}
at the center of the sphere. These regular  solutions  must be computed numerically. In the case of bounded isothermal systems, we must stop the integration at the normalized box radius 
\begin{equation}
\alpha=(4\pi G\beta m\rho_{0})^{1/2}R.
\label{E5bis}
\end{equation}

We shall now express the parameter $\alpha$ in terms of the
temperature and the energy.  According to the Poisson equation (\ref{R5}),
we have
\begin{equation}
GM=\int_{0}^{R} 4\pi G\rho r^{2}dr=\int_{0}^{R}{d\over dr}\biggl (r^{2}{d\Phi\over dr}\biggr )dr=\biggl (r^{2}{d\Phi\over dr}\biggr )_{r=R},
\label{E6}
\end{equation}
which is just a particular case of the Gauss theorem.
Introducing the dimensionless variables defined previously, we get
\begin{equation}
\eta\equiv {\beta GMm\over R}=\alpha\psi'(\alpha).
\label{E7}
\end{equation}
The relation between  $\alpha$ and the normalized temperature $\eta$ is not affected by special relativity.

For the total energy, using the Virial theorem (\ref{V9}) and the expression (\ref{R16}) for the kinetic term, we have
\begin{equation}
E=K+W={\cal F}(x)Mc^{2}-{3N\over\beta}+3p(R)V.
\label{E8}
\end{equation}
Now, the pressure at the boundary of the domain can be written
\begin{equation}
p(R)={\rho(R)\over m\beta}={\rho_{0}e^{-\psi(\alpha)}\over m\beta}.
\label{E9}
\end{equation}
Expressing the central density in terms of $\alpha$, using Eq. (\ref{E5bis}), we have equivalently
\begin{equation}
p(R)={\alpha^{2}\over 4\pi G\beta^{2}m^{2}R^{2}}e^{-\psi(\alpha)}.
\label{E11}
\end{equation}
The total energy therefore reads
\begin{equation}
\Lambda\equiv -{ER\over GM^{2}}=-{Rc^{2}\over GM}{\cal F}(x)+{3\over\alpha\psi'(\alpha)}-{e^{-\psi(\alpha)}\over\psi'(\alpha)^{2}},
\label{E12}
\end{equation}
where we have used Eq. (\ref{E7}) to eliminate the temperature in the last two terms. 

It will be convenient in the following to introduce the parameter
\begin{equation}
\mu={Rc^{2}\over GM}\equiv {2R\over R_{S}^{class}},
\label{E13}
\end{equation}
which is twice the ratio between the system radius $R$ and the ``classical'' Schwarzschild radius
\begin{equation}
R_{S}^{class}={2GM\over c^{2}}.
\label{E14}
\end{equation}
Clearly, our semi-relativistic treatment (ignoring general relativity) is only valid for $\mu\gg 1$. However, in our rather formal analysis, we shall treat $\mu$ as a free parameter varying in the range $0\le\mu<+\infty$. The relativistic parameters $\mu$ and $x$ are related by
\begin{equation}
x=\mu\eta.
\label{E15}
\end{equation}
Therefore, in accordance with Eq. (\ref{E12}) and (\ref{E15}),  the relation between the parameter $\alpha$ and the normalized energy $\Lambda$ takes 
the form 
\begin{equation}
\Lambda\equiv -{ER\over GM^{2}}=-\mu {\cal F}\lbrack \mu\alpha\psi'(\alpha)\rbrack +{3\over\alpha\psi'(\alpha)}-{e^{-\psi(\alpha)}\over\psi'(\alpha)^{2}}.
\label{E16}
\end{equation}
In the classical limit ($\mu\rightarrow +\infty$), we recover the result of Lynden-Bell \& Wood (1968)
\begin{equation}
\Lambda={3\over 2\alpha\psi'(\alpha)}-{e^{-\psi(\alpha)}\over\psi'(\alpha)^{2}},
\label{E17}
\end{equation}
and in the formal limit $\mu\rightarrow 0$, we get
\begin{equation}
\Lambda=-{e^{-\psi(\alpha)}\over\psi'(\alpha)^{2}}.
\label{E18}
\end{equation}

\begin{figure}[htbp]
\centerline{
\psfig{figure=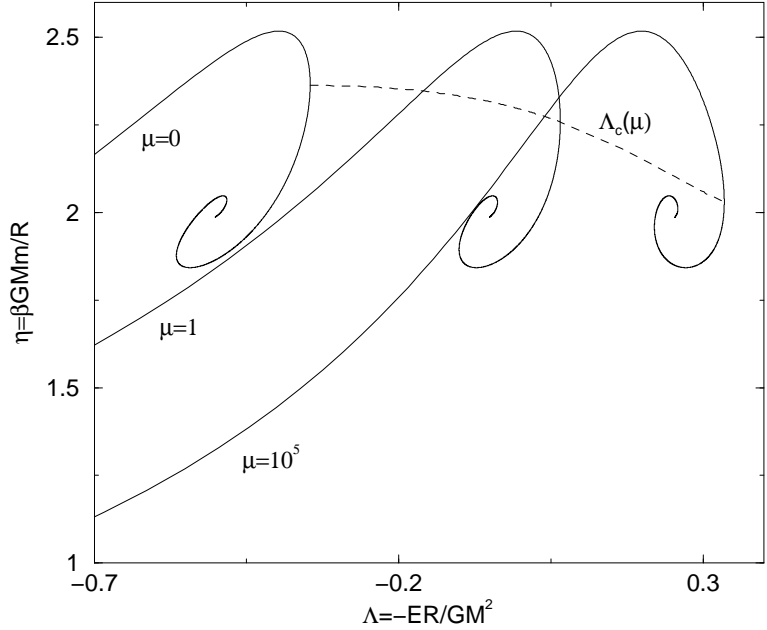,angle=0,height=8.5cm}}
\caption{Equilibrium phase diagram for isothermal gas spheres described by a relativistic equation of state. Relativistic effects shift the onset of instability to larger energies.}
\label{etalambdamu}
\end{figure}

\begin{figure}[htbp]
\centerline{
\psfig{figure=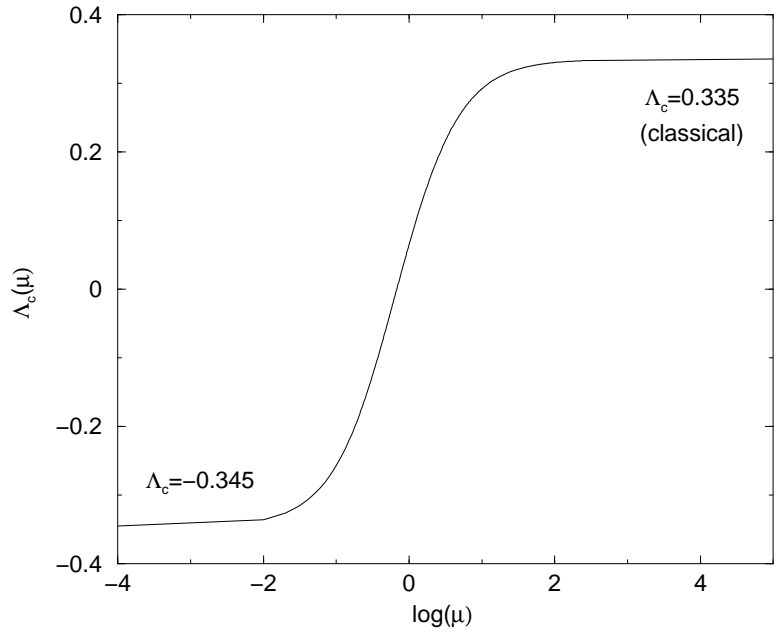,angle=0,height=8.5cm}}
\caption{Critical energy $\Lambda_{c}$ as a function of the relativistic parameter $\mu$. }
\label{lambdacmu}
\end{figure}

On Fig. \ref{etalambdamu}, we have represented the equilibrium phase
diagram $\beta-E$ for different values of the relativistic parameter
$\mu$. We see that the effect of relativity is to shift the spiral to
the left. Therefore, the ``gravothermal catastrophe'' corresponding to
the absence of equilibrium below a critical energy occurs {\it sooner}
than in the Newtonian case. The critical energy is ploted as a
function of the relativistic parameter $\mu$ on
Fig. \ref{lambdacmu}. For $\mu\rightarrow +\infty$, we recover the
classical result of Antonov $\Lambda_{c}(+\infty)=0.335$ and, in the
formal limit $\mu\rightarrow 0$, we get
$\Lambda_{c}(0)=-0.345$. Clearly, the spiral is not destroyed by
relativistic effects.  Only is its shape slightly modified: the
relativistic spirals are more ``stretched'' than the classical
one. Note that the maximum value of $\eta$ is independant on $\mu$ and
is equal to its classical value $\eta_{c}=2.52$. On the other hand,
substituting the expression (\ref{E4}) for the singular sphere in
Eqs. (\ref{E7})-(\ref{E16}), we find that the center of the spiral is
defined by the equations
\begin{equation}
\eta_{s}=2,
\label{E20}
\end{equation} 
\begin{equation}
\Lambda_{s}(\mu)=1-\mu{\cal F}(2\mu).
\label{E21}
\end{equation} 
In the limit $\mu\rightarrow +\infty$, $\Lambda_{s}(+\infty)={1\over
4}$ and in the limit $\mu\rightarrow 0$, $\Lambda_{s}(0)=-{1\over
2}$. 

\begin{figure}[htbp]
\centerline{
\psfig{figure=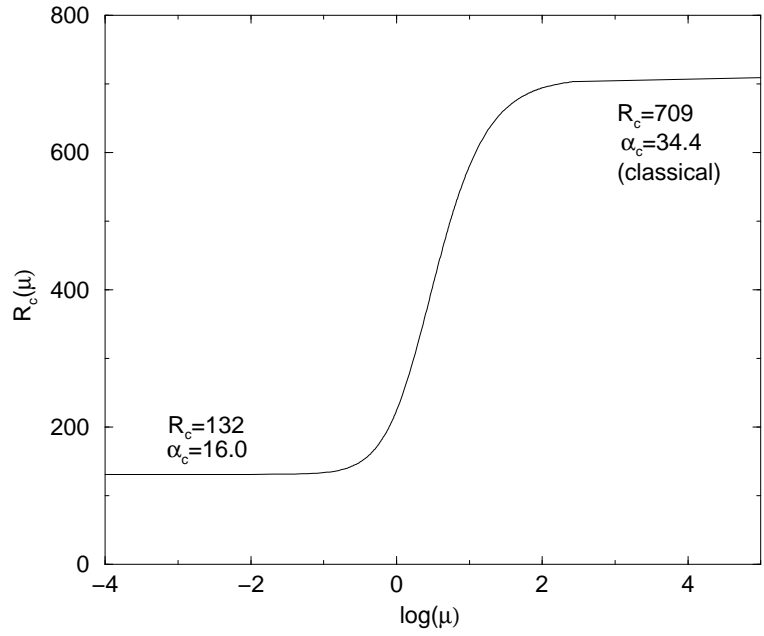,angle=0,height=8.5cm}}
\caption{Critical density contrast ${\cal R}_{c}$ as a function of the relativistic parameter $\mu$. }
\label{contrastemu}
\end{figure}

The spiral is parametrized by the normalized box radius $\alpha$
that goes from $0$ (ordinary gas) to $+\infty$ (singular sphere) when
we spiral inwards. If one prefers, we can use a parametrization in
terms of the density contrast
\begin{equation}
{\cal R}={\rho_{0}\over \rho(R)}=e^{\psi(\alpha)},
\label{E19}
\end{equation}
that goes from $1$ to $+\infty$. On Fig. \ref{contrastemu}, we plot
the critical density contrast (corresponding to $\Lambda_{c}$) as a
function of the relativistic parameter $\mu$. For $\mu\rightarrow
+\infty$, we recover the classical value ${\cal R}_{c}(+\infty)=709$
(and $\alpha_{c}(+\infty)=34.4$). For $\mu=0$, we get ${\cal
R}_{c}(0)=132$ (and $\alpha_{c}(0)=16.0$). It is found that
instability occurs for smaller density contrasts when relativity is accounted for. 

\subsection{The Milne variables}
\label{sec_milne}

\begin{figure}[htbp]
\centerline{
\psfig{figure=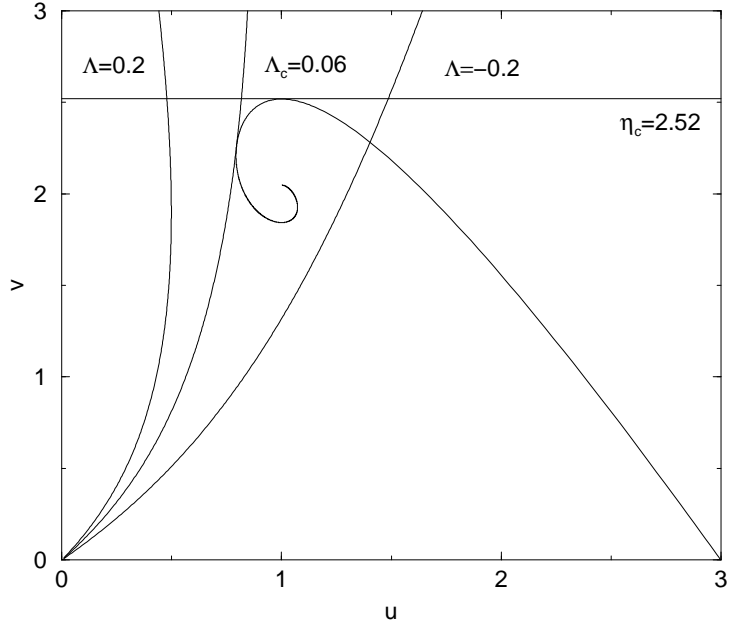,angle=0,height=8.5cm}}
\caption{The $(u,v)$ plane. All isothermal spheres must necessarily lie on the spiral. There exists solutions in the canonical ensemble only for $\eta<2.52$. In the microcanonical ensemble, the critical energy depends on the relativistic parameter $\mu$. For $\mu=1$, there exists solutions only for $\Lambda<\Lambda_{c}=0.06$ }
\label{Lcrituv1}
\end{figure}

It will be convenient in the following to introduce the Milne variables $(u,v)$ defined by (Chandrasekhar 1942)
\begin{equation}
u={\xi e^{-\psi}\over\psi'},\qquad {\rm and}\qquad v=\xi\psi'.
\label{M1}
\end{equation}
Taking the logarithmic derivative of $u$ and $v$ with respect to $\xi$ and using Eq. (\ref{E3}), we get
\begin{equation}
{1\over u}{du\over d\xi}={1\over\xi}(3-v-u),
\label{M2}
\end{equation}
\begin{equation}
{1\over v}{dv\over d\xi}={1\over\xi}(u-1).
\label{M3}
\end{equation}
Taking the ratio of these equations, we find that the variables $u$ and $v$ are related to each other by a first order differential equation 
\begin{equation}
{u\over v}{dv\over du}=-{u-1\over u+v-3}.
\label{M4}
\end{equation}
The solution curve in the $(u,v)$ plane is well-known and is represented on Fig. \ref{Lcrituv1}. Its striking oscillating behavior has been described by a number of authors (see in particular Chandrasekhar 1942). We refer to Padmanbhan (1989) and Chavanis (2001) for the description of its main characteristics in connexion with the present work.

\begin{figure}[htbp]
\centerline{
\psfig{figure=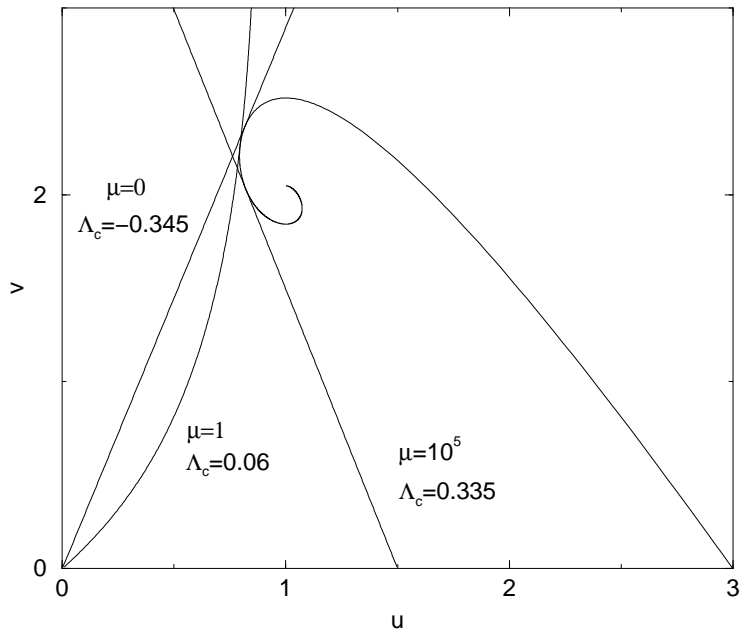,angle=0,height=8.5cm}}
\caption{Same as Fig. \ref{Lcrituv1} but for different values of $\mu$ and, in each case, for the critical parameter $\Lambda_{c}(\mu)$ above which there is no equilibrium solution.}
\label{Lcrituv2}
\end{figure}

It turns out that the normalized temperature and the normalized energy can be expressed very simply in terms of the values of $u$ and $v$ at the normalized box radius $\alpha$. Indeed, writing $u_{0}=u(\alpha)$ and $v_{0}=v(\alpha)$ and using Eqs. (\ref{E7}) (\ref{E16}), we get
\begin{equation}
\eta=v_{0},
\label{M5}
\end{equation}
\begin{equation}
\Lambda=-\mu{\cal F}(\mu v_{0})+{3\over v_{0}}-{u_{0}\over v_{0}}.
\label{M6}
\end{equation}
The intersections between the curves defined by Eq. (\ref{M5})-(\ref{M6}) and the spiral in the $(u,v)$ plane determine the values of $\alpha$ corresponding to a given temperature or energy. Considering Eq. (\ref{M5}), we find that there is no intersection for $\eta={\beta GM\over R}>v_{max}=2.52$. In the canonical ensemble, a gaseous sphere is expected to collapse below a critical temperature $kT_{c}={GmM\over 2.52 R}$. This classical result is not altered by special relativity. Considering now the microcanonical ensemble, we first note that Eq. (\ref{M6}) can be rewritten
\begin{equation}
u_{0}=3-\mu v_{0}{\cal F}(\mu v_{0})-\Lambda v_{0}.
\label{M7}
\end{equation}
In the classical limit ($\mu\rightarrow +\infty$) it reduces to the straight
line found by Padmanbhan (1989)
\begin{equation}
u_{0}={3\over 2}-\Lambda v_{0},
\label{M8}
\end{equation}
and in the limit $\mu\rightarrow 0$, we find another straight line
\begin{equation}
u_{0}=-\Lambda v_{0}.
\label{M9}
\end{equation}
The curve (\ref{M7}) is ploted on Fig. \ref{Lcrituv1} for a fixed
value of $\mu$ and for different values of $\Lambda$. For
$\Lambda>\Lambda_{c}(\mu)$ there is no intersection, for
$\Lambda=\Lambda_{c}(\mu)$ the curve (\ref{M7}) is tangent to the
spiral and for $\Lambda=\Lambda_{c}(\mu)$ there are one or several
intersections. We recover therefore by this graphical construction the
existence of a critical energy below which no hydrostatic equilibrium
can exist for isothermal spheres.  On Fig. \ref{Lcrituv2}, we plot the
same diagram as Fig. \ref{Lcrituv1} but for different values of $\mu$
and, in each case, for the critical energy $\Lambda_{c}(\mu)$. It
confirms that instability occurs sooner when relativistic effects are
taken into account.

\subsection{The condition of thermodynamical stability}
\label{sec_cond}

We now address the question of thermodynamical stability. Let us recall that the isothermal spheres lying on the spiral are critical points of entropy but they are not necessarily entropy {\it maxima}. To determine whether they are local entropy maxima or saddle points, we must examine the sign of the second variations of entropy given by Eq. (\ref{F11}). In fact, this condition of stability has the same form as in the classical case provided that we make the substitution 
\begin{equation}
 -{1\over 3NkT^{2}}\  \rightarrow \ {k\over 2Mmc^{4}}{1\over {\cal F}'(x)}.
\label{C1}
\end{equation}
Therefore, the analysis of Padmanbhan (1989) for the classical Antonov instability can be extended straightforwardly. Introducing the mass perturbation $q(r)\equiv \delta M(r)=\int_{0}^{r}4\pi r^{'2}\delta\rho(r')dr'$ within the sphere of radius $r$ such that
\begin{equation}
\delta\rho={1\over 4\pi r^{2}}{dq\over dr},
\label{C2}
\end{equation}
the second variations of entropy can be put in a quadratic form
\begin{equation}
\delta^{2}S=\int_{0}^{R}\int_{0}^{R}drdr' q(r)K(r,r')q(r'),
\label{C3}
\end{equation}
with
\begin{equation}
K(r,r')={k\over 2Mmc^{4}}{1\over {\cal F}'(x)}{d\Phi\over dr}(r){d\Phi\over dr}(r')+{1\over 2}\delta(r-r')\biggl \lbrack {G\over Tr^{2}}+{k\over m}{d\over dr}\biggl ({1\over 4\pi\rho r^{2}}{d\over dr}\biggr )\biggr\rbrack.
\label{C4}
\end{equation}
Clearly, the conservation of mass imposes the boundary conditions $q(0)=q(R)=0$. The problem of stability can therefore be reduced to the study of the eigenvalue equation
\begin{equation}
\int_{0}^{R}K(r,r')F_{\lambda}(r')dr'=\lambda F_{\lambda}(r),
\label{C5}
\end{equation}
with the boundary conditions $F_{\lambda}(0)=F_{\lambda}(R)=0$.  If all
the eigenvalues are negative, then $\delta^{2}S<0$ and the critical point
is a local entropy maximum. If one eigenvalue is positive, the
critical point is an unstable saddle point. The point of marginal
stability $\Lambda_{c}$ is determined by the condition that the
largest eigenvalue is equal to zero ($\lambda=0$). We thus have to
solve the differential equation
\begin{equation}
\biggl \lbrack {k\over m}{d\over dr}\biggl ({1\over 4\pi\rho r^{2}}{d\over dr}\biggr )+{G\over Tr^{2}}\biggr \rbrack F(r)=-{k\over Mmc^{4}}{1\over {\cal F}'(x)}V{d\Phi\over dr}(r),
\label{C6}
\end{equation}
with
\begin{equation}
V=\int_{0}^{R}{d\Phi\over dr}(r')F(r')dr',
\label{C7}
\end{equation}
and $F(0)=F(R)=0$. Introducing the dimensionless variables defined in section \ref{sec_equi}, it can be rewritten
\begin{equation}
\biggl\lbrack {d\over d\xi}\biggl ({e^{\psi}\over\xi^{2}}{d\over d\xi}\biggr )+{1\over \xi^{2}}\biggr\rbrack F(\xi)=\chi {d\psi\over d\xi},
\label{C8}
\end{equation}
with
\begin{equation}
\chi=-{1\over x^{2}{\cal F}'(x)} {1\over \alpha^{2}\psi'(\alpha)}\int_{0}^{\alpha}{d\psi\over d\xi}(\xi')F(\xi')d\xi',
\label{C9}
\end{equation}
and $F(0)=F(\alpha)=0$. As shown by Padmanabhan (1989), the solutions of the differential equation (\ref{C8}) can be expressed in terms of the solutions of the Emden equation (\ref{E3}) as
\begin{equation}
F(\xi)=\chi\biggl\lbrack {1\over 1-u_{0}}(\xi^{3}e^{-\psi}-\xi^{2}\psi')+\xi^{2}\psi'\biggr\rbrack.
\label{C10}
\end{equation}
We can check that this function satisfies the boundary conditions $F(0)=F(\alpha)=0$. The point of marginal stability is obtained  by substituting the solution (\ref{C10}) in Eq. (\ref{C9}). The integrations can be carried out and the solutions expressed in terms of the Milne variables $u_{0}$ and $v_{0}$ (see Padmanabhan (1989) for more details). In our semi-relativistic treatment, we obtain
\begin{equation}
2u_{0}^{2}+u_{0}v_{0}-7u_{0}+3-\mu^{2}v_{0}^{2}{\cal F}'(\mu v_{0})(u_{0}-1)=0,
\label{C11}
\end{equation}
where we have used Eqs. (\ref{E15}) and (\ref{M5}) to simplify the last term. 

In the classical limit $\mu\rightarrow +\infty$ we recover the result of Padmanabhan (1989)
\begin{equation}
4u_{0}^{2}+2u_{0}v_{0}-11u_{0}+3=0,
\label{C12}
\end{equation}
and in the formal limit $\mu\rightarrow 0$, we find
\begin{equation}
2u_{0}+v_{0}-4=0.
\label{C13}
\end{equation}

The intersections between the curve (\ref{C11}) and the spiral in the $(u,v)$ plane determine the values of $\alpha$ for which a new mode of stability is lost (i.e., a new eigenvalue $\lambda$ becomes positive). Since the curve (\ref{C11}) passes through the singular sphere $(u_{s},v_{s})=(1,2)$ at the center of the spiral, there is an infinity of intersections. The first intersection (for which $\alpha$ is minimum) corresponds to the point of marginal stability. We can show that the points determined by Eq. (\ref{C11}) are precisely those for which $\Lambda$ is extremum in agreement with the turning point analysis of Katz (1978). Indeed, differentiating the expression (\ref{M6}) for $\Lambda$ with respect to $\alpha$, we get
\begin{equation}
{d\Lambda\over d\alpha}=-\mu^{2}{\cal F}'(\mu v_{0}){dv_{0}\over d\alpha}-{3\over v_{0}^{2}}{dv_{0}\over d\alpha}-{1\over v_{0}}{du_{0}\over d\alpha}+{u_{0}\over v_{0}^{2}}{dv_{0}\over d\alpha}.
\label{C14}
\end{equation}
Using equations (\ref{M2})-(\ref{M3}), we obtain
\begin{equation}
{d\Lambda\over d\alpha}={1\over \alpha v_{0}}\biggl\lbrack 2u_{0}^{2}+u_{0}v_0-7u_0+3-\mu^{2}v_0^{2}{\cal F}'(\mu v_0)(u_0 -1)\biggr \rbrack,
\label{C15}
\end{equation}
and we check that the condition ${d\Lambda\over d\alpha}=0$ is equivalent to Eq. (\ref{C11}).

\begin{figure}[htbp]
\centerline{
\psfig{figure=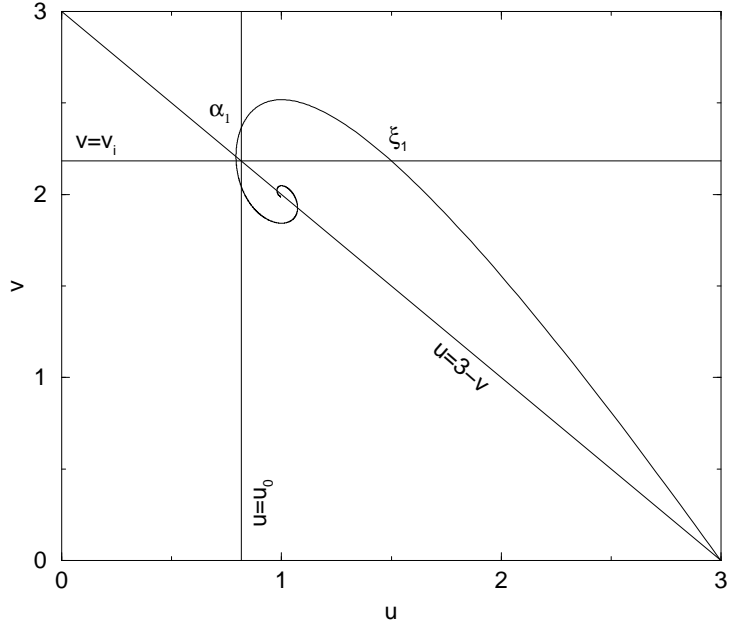,angle=0,height=8.5cm}}
\caption{Graphical construction to determine the nodes of the perturbation profile $\delta\rho$ at the point of marginal stability. The construction is done explicitly for $\mu=0$, for which $\alpha_{1}=16.0$ and $u_{0}=0.817$. There is only one zero satisfying $\xi_{1}<\alpha_{1}$ so that the perturbation profile does not present a ``core-halo'' structure. This property is maintained until $\mu>1.61$. }
\label{nodeuv}
\end{figure}

\begin{figure}[htbp]
\centerline{
\psfig{figure=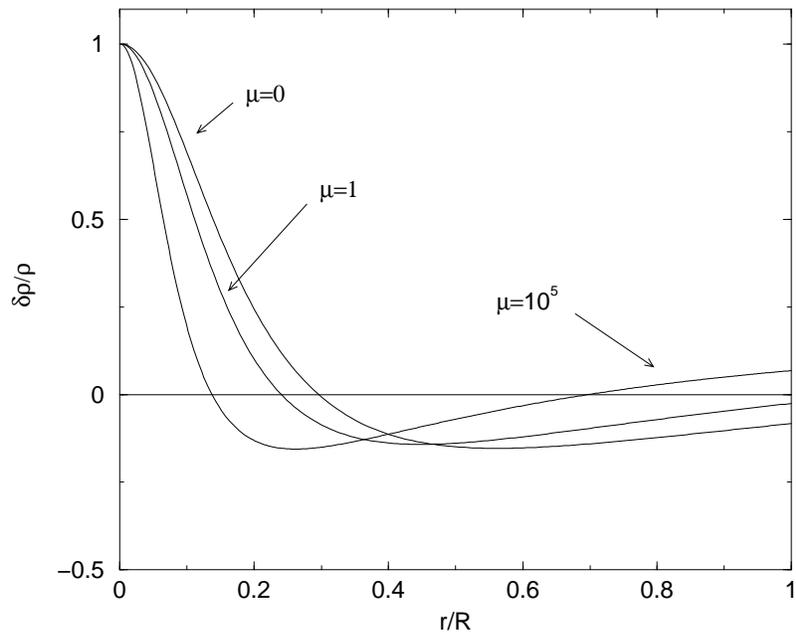,angle=0,height=8.5cm}}
\caption{Density perturbation profile at the point of marginal stability as a function of the relativistic parameter. }
\label{deltarho}
\end{figure}

It is also easy to determine the form of the perturbation that triggers the instability at the critical point. According to Eq. (\ref{C2}), the eigenfunction associated with the eigenvalue $\lambda=0$ can be written
\begin{equation}
{\delta\rho\over\rho_{0}}={1\over 4\pi\xi^{2}}{dF\over d\xi},
\label{C16}
\end{equation}
where $F(\xi)$ is given by Eq. (\ref{C10}). Simplifying the derivative with the aid of Eq. (\ref{E3}), we can express the perturbation profile in terms of the Milne variables (\ref{M1}) as 
\begin{equation}
{\delta\rho\over\rho}={\chi\over 4\pi}{1\over 1-u_{0}}(3-v-u_{0}).
\label{C17}
\end{equation}
The qualitative behavior of the perturbation profile can be studied without numerical integration by a graphical construction (see Padmanabhan 1989). The density perturbation $\delta\rho$ becomes zero at the point(s) $\xi_{i}$ such that $u_{0}=3-v(\xi_{i})$.  On Fig. \ref{nodeuv}, we first draw the line $u=3-v$. This lines passes through the singular sphere $(u_{s}=1,v_{s}=2)$ and also cuts the spiral at the points of vertical tangent (see Eq. (\ref{M2})). In particular, the first intersection corresponds to $\alpha_{*}=22.5$ and $(u_{*},v_{*})=(0.793,2.208)$. Then, we draw the line $u=u_{0}=u(\alpha_{1})$. The intersection between these two lines determines $v(\xi_{i})$. The intersection between $v=v(\xi_{i})$ and the spiral determines the zeros of $\delta\rho$. For $\alpha>\alpha_{*}$, there are two intersections satisfying $\xi_{i}<\alpha_{1}$ so that the perturbation profile presents a ``core-halo'' structure. This is the case in particular in the classical limit $\mu\rightarrow +\infty$ for which $\alpha_{1}=34.4$ (see Padmanabhan 1989). By contrast, for $\alpha<\alpha_{*}$, there is only one intersection satisfying $\xi_{1}<\alpha_{1}$ so that the perturbation profile does {\it not} present a ``core-halo'' structure. This is the case in particular in the (formal) limit $\mu=0$ for which $\alpha_{1}=16.0$. The ``core-halo'' structure disappears for $\alpha_{1}=\alpha_{*}=22.5$ corresponding to a relativistic parameter $\mu_{*}=1.61$. Since our study is valid for $\mu\gg 1$, we deduce that the density profile always presents a ``core-halo'' structure in the cases of physical interest. However, relativistic effects have the tendency to reduce the extent of the halo (see Fig. \ref{deltarho}).

The previous results are valid in the microcanonical ensemble in which the energy is fixed. In the canonical ensemble, we must consider maxima of the free energy $J=S-\beta E$ at fixed temperature. In that case, the condition of stability is given by Eq. (\ref{C6}) with $V=0$. Since the relativistic function ${\cal F}(x)$ does not appear anymore in the equations, we conclude that special relativity does not change the classical results in the canonical ensemble. In particular, the perturbation profile does not present a ``core-halo'' structure at the critical point (Chavanis 2001).

\section{Isothermal gaseous spheres in general relativity}
\label{sec_iso}

\subsection{The equations governing equilibrium}
\label{sec_eq}

We now address the structure and the stability of isothermal gas spheres in the context of general relativity. The Einstein field equations of general relativity are expressed as  
\begin{equation}
R_{\mu\nu}-{1\over 2}g_{\mu\nu}R=-{8\pi G\over c^{4}} T_{\mu\nu},
\label{eq1}
\end{equation}
where $R_{\mu\nu}$ is the Ricci tensor, $T_{\mu\nu}$ the energy-momentum tensor and $g_{\mu\nu}$ the metric tensor defined by
\begin{equation}
ds^{2}=-g_{\mu\nu}dx^{\mu}dx^{\nu},
\label{eq2}
\end{equation}
where $ds$ is the invariant interval between two neighbouring space-time events. 

In the following, we shall restrict ourselves to spherically symmetric systems with motions, if any, only in the radial directions. Under these assumptions, the metric can be written in the form
\begin{equation}
ds^{2}=e^{\nu}d\tau^{2}-r^{2}(d\theta^{2}+\sin^{2}\theta d\phi^{2})-e^{\lambda}dr^{2}, \quad \tau=ct,
\label{eq3}
\end{equation}
where $\nu$ and $\lambda$ are functions of $r$ and $\tau$ only. The energy-momentum tensor is assumed to be that for a perfect fluid
\begin{equation}
T^{\mu\nu}=p g^{\mu\nu}+(p+\epsilon)u^{\mu}u^{\nu},
\label{eq4}
\end{equation}
where $u^{\mu}=dx^{\mu}/ds$ is the fluid four-velocity, $p$ is the isotropic 
pressure and $\epsilon$ is the energy density including the rest mass.

The equations of general relativity governing the hydrostatic equilibrium of a spherical distribution of matter are well known. They are given by (see, e.g., Weinberg 1972)
\begin{equation}
{d\over dr}(re^{-\lambda})=1-{8\pi G\over c^{4}}r^{2}\epsilon,
\label{eq5}
\end{equation}
\begin{equation}
{dp\over dr}=-{1\over 2}(\epsilon+p){d\nu\over dr},
\label{eq6}
\end{equation}
\begin{equation}
{e^{-\lambda}\over r}{d\nu\over dr}={1\over r^{2}}(1-e^{-\lambda})+{8\pi G\over c^{4}}p.
\label{eq7}
\end{equation}
These equations can be deduced from the Einstein equations (\ref{eq1}). However, Eq. (\ref{eq6}) can be obtained more directly from the local law of energy-momentum conservation, $D_{\mu}T^{\mu\nu}=0$, which is also contained in the Einstein equations.

Equations (\ref{eq5})-(\ref{eq7}) can be combined to give
\begin{equation}
\biggl \lbrace 1-{2GM(r)\over c^{2}r}\biggr \rbrace {dp\over dr}=-{1\over c^{2}}(\epsilon+p)\biggl\lbrace {GM(r)\over r^{2}}+{4\pi G\over c^{2}}pr\biggr\rbrace,
\label{eq8}
\end{equation}
and
\begin{equation}
M(r)={4\pi\over c^{2}}\int_{0}^{r}\epsilon r^{2}dr.
\label{eq9}
\end{equation}
These equations are known as the Oppenheimer-Volkoff (1939) equations. They extend the classical condition of hydrostatic equilibrium for a star in general relativity. Using equations (\ref{eq5}) and (\ref{eq7}), we find that the functions $\lambda(r)$ and $\nu(r)$ satisfy the relations
\begin{equation}
e^{-\lambda}=1-{2G\over rc^{2}}M(r),
\label{eq10}
\end{equation}
and
\begin{equation}
{d\nu\over dr}={1+4\pi p r^{3}/M(r)c^{2}\over r  ({rc^{2}/ 2 GM(r)}-1 )}.
\label{eq11}
\end{equation}
In the empty space outside the star, $p=\epsilon=0$. Therefore, if $M=M(R)$ denotes the total mass-energy of the star, Eqs. (\ref{eq10})(\ref{eq11}) become for $r>R$
\begin{equation}
e^{-\lambda}=1-{2GM\over rc^{2}}, \qquad {d\nu\over dr}={1\over r  ({rc^{2}/ 2 GM}-1 )}.
\label{eq12}
\end{equation}
The second equation is readily integrated in 
\begin{equation}
\nu=\ln(1-2GM/rc^{2}).
\label{eq13}
\end{equation}
Substituting the foregoing expressions for $\lambda$ and $\nu$ in Eq. (\ref{eq3}), we obtain the well-known Schwarzschild's form of the metric outside a star
\begin{equation}
ds^{2}=\biggl (1-{2GM\over rc^{2}}\biggr )d\tau^{2}-r^{2}(d\theta^{2}+\sin^{2}\theta d\phi^{2})-{dr^{2}\over 1-{2GM/rc^{2}}}.
\label{eq14}
\end{equation}
This metric is singular at
\begin{equation}
r={2GM\over c^{2}}\equiv R_{S},
\label{eq15}
\end{equation}
where $R_{S}$ is the Schwazschild radius appropriate to the mass $M$. This does not mean that spacetime is singular at that radius but only that this particular metric is. Indeed, the singularity can be removed by a judicious change of coordinate system (see, e.g., Weinberg 1972). When $R_{S}>R$, the star is a black hole and no particle or even light can leave the region $R<r<R_{S}$. However, in our case the discussion does not arise because $R_{S}<R$. Indeed, for a gaseous sphere in hydrostatic equilibrium, it can be shown that the radius of the configuration is necessarily restricted by the inequality (Buchdahl 1959)
\begin{equation}
R\ge {9\over 8}\ {2GM\over c^{2}}={9\over 8}R_{S}.
\label{eq16}
\end{equation}
Therefore, the points exterior to the star always satisfy $r>R_{S}$.

\subsection{The equation of state}
\label{sec_state}

To close the system of equations (\ref{eq8})-(\ref{eq9}), we need to
specify an equation of state relating the pressure $p$ to the energy
density $\epsilon$. Quite generally, the first law of thermodynamics
can be expressed as
\begin{equation}
d\biggl ({\epsilon\over n}\biggr )=-pd\biggl ({1\over n}\biggr )+Tds,
\label{S1}
\end{equation}
or, equivalently,
\begin{equation}
d\epsilon={p+\epsilon\over n}dn+nTds,
\label{S2}
\end{equation}
where $n$ is the baryon number density and $s$ the entropy per baryon in rest frame. It must be completed by two equations of state $p=p(n,s)$ and $T=T(n,s)$. Then, Eq. (\ref{S2}) can be integrated to give $\epsilon(n,s)$ (see, e.g., Misner {\it et al.} 1973).

We shall assume in the following that the term $Tds$ in Eq. (\ref{S2}) can be neglected. This simplification arises in two different situations. In the case of neutron stars or white dwarfs, the thermal energy $kT$ is much smaller than the Fermi energy, so the neutrons or the electrons are completely degenerate ($kT\ll E_{fermi}$). On the other hand, in supermassive stars, temperature and entropy are important but convection keeps the star stired up and produces a uniform entropy distribution ($s={\it Cst.}$). In these two important situations, the first law of thermodynamics reduces to
\begin{equation}
d\epsilon={p+\epsilon\over n}dn,
\label{S3}
\end{equation}
and we just require one equation of state $p=p(n)$. We shall consider an equation of state of the form
\begin{equation}
p=K n^{\gamma},
\label{S4}
\end{equation}
where $K$ and $\gamma$ are constant. It is easy to check that the general solution of Eq. (\ref{S3}) with Eq. (\ref{S4}) is
\begin{equation}
\epsilon=A p^{1/\gamma}+{1\over \gamma-1}p,
\label{S5}
\end{equation}
where $A$ is a constant. Systems obeying the pressure-energy density relation (\ref{S5}) with $A\neq 0$ have been considered by Tooper (1965). Since the relation between $\epsilon$ and $p$ is essentially a power-law, these systems generalize the polytropes of Newtonian theory. We shall here consider the limiting situation  $A=0$. In that case, Eq. (\ref{S5}) reduces to the so-called ``gamma law'' equation of state
\begin{equation}
p=q\epsilon\qquad {\rm with}\qquad q=\gamma-1.
\label{S6}
\end{equation}
Since the relation between $p$ and $\epsilon$ is linear, this equation of state extends the theory of isothermal spheres to the context of general relativity. As noted by Chandrasekhar (1972), for this equation of state, the Oppenheimer-Volkoff equations become mathematically similar to those describing a classical isothermal gas (the so-called Emden equations). 

An equation of state of the form (\ref{S6}) has been introduced in different situations: 

(i) This equation of state prevails in the highly energetic core of neutron stars where the matter is ultra-relativistic  and completely degenerate (Oppenheimer \& Volkoff 1939, Misner \& Zapolsky 1964, Meltzer \& Thorne 1966). If the system is modelled as a pure collection of noninteracting fermions, standard theory leads to (Chandraskhar 1942)
\begin{equation}
p={1\over 3}\epsilon \qquad  (\gamma={4/3} ).
\label{S7}
\end{equation} 
In that case, the constant $K$ which appears in Eq. (\ref{S4}) is explicitly given by $K={1\over 8}({3\over\pi})^{1/3}hc$, where $h$ is the Planck constant. Other versions of the equation of state have attempted to take into account nucleon-nucleon interactions or a spectrum of baryon species. Equations of the form (\ref{S6}) but with numerical coefficient other than $1/3$ have sometimes been suggested. For example Ambatsumyan \& Saakyan (1961) have used an equation with $q=1/13$ ($\gamma=14/13$) and Tsuruta \& Cameron (1966) with $q=1$. The equation of state
\begin{equation}
p=\epsilon,
\label{S8}
\end{equation}  
was also introduced by Zel'dovich (1962) for a gas of baryons interacting through a vector meson field. This represents an extremely high pressure since the sound velocity, $(dp/d\epsilon)^{1/2}c$, is equal to the velocity of light for this value of $q$. This is clearly an upper bound. In the following, we shall consider a general  equation of state of the form (\ref{S6}) with $q$ in the range $0\le q\le 1$.

(ii) Bisnovatyi-Kogan \& Zel'dovich (1969) and Bisnovatyi-Kogan \& Thorne (1970) have considered the general relativistic equilibrium configuration of a gas whose temperature $T$ is constant everywhere. In that case, the equation of state is  given by (see section \ref{sec_antonov}) 
\begin{equation}
p={K_{2}({mc^{2}\over kT})\over {mc^{2}\over kT}K_{3}({mc^{2}\over kT})-K_{2}({mc^{2}\over kT})}\epsilon,
\label{S9}
\end{equation}
which  is again of the form (\ref{S6}). In particular, $q={1\over 3}$ when $T\rightarrow +\infty$. However, as Bisnovatyi-Kogan \& Thorne emphasize, the equation of state (\ref{S9}) can only describe a ``local'' thermodynamical equilibrium. Indeed, ``global'' thermodynamical equilibrium in general relativity requires that the {\it redshifted temperature} $e^{\nu(r)/2}T$ be uniform throughout the medium (Tolman 1934). Thus, the spheres described by Eq. (\ref{S9}) are not ``isothermal'' in the general-relativistic sense; heat will slowly diffuse inward in them, upsetting the condition $T={\it Cst.}$ and trying to establish $e^{\nu(r)/2}T(r)={\it Cst.}$ instead. However, since the hydrostatic equations  (\ref{eq8})(\ref{eq9}) with the equation of state (\ref{S6}) are mathematically similar to the classical Emden equation for isothermal gas spheres, we shall call the equation of state (\ref{S6}) ``isothermal'', following the terminology of Chandrasekhar (1972), although this is only correct in a local sense.

(iii) The equation of state (\ref{S6}) was also proposed by Saslaw {\it et al.} (1996) in a cosmological context to model the ultimate state of an Einstein-de Sitter universe that undergoes a phase transition caused by gravitational clustering. This phase transition can lead to the growth of a centrally concentrated distribution of matter so that the universe would pass from a statistically homogeneous state to a state of rotational symmetry around one point only. The resulting configuration neither expands nor contracts, so the global solution is stationary. According to Saslaw {\it et al.} (1996), this ``isothermal universe'' would represent the ultimate astrophysical prediction.

(iv) Note finally that the equation of state $p={1\over 3}\epsilon$ prevails  in a medium where the pressure is entirely due to radiation (Chandrasekhar 1942).

\subsection{The general relativistic Emden equation}
\label{sec_emden}

Considering the equation of state (\ref{S6}), we shall introduce the dimensionless variables $\xi$, $\psi$ and $M(\xi)$ by the relations
\begin{equation}
\epsilon=\epsilon_{0}e^{-\psi},\qquad\qquad r=\biggl \lbrace {c^{4}q\over 4\pi G\epsilon_{0}(1+q)}\biggr\rbrace^{1/2}\xi,
\label{em1}
\end{equation}
and 
\begin{equation}
M(r)={4\pi\epsilon_{0}\over c^{2}}\biggl\lbrace {c^{4}q\over 4\pi G\epsilon_{0}(1+q)}\biggr\rbrace^{3/2}M(\xi).
\label{em2}
\end{equation}

In terms of the variables $\psi$ and $\xi$, Eqs. (\ref{eq8}) and (\ref{eq9}) can be reduced to the following dimensionless forms (Chandrasekhar 1972)
\begin{equation}
\biggl\lbrace 1-{2q\over 1+q}{M(\xi)\over\xi}\biggr\rbrace {d\psi\over d\xi}={M(\xi)\over\xi^{2}}+q\xi e^{-\psi},
\label{em3}
\end{equation}
and
\begin{equation}
{dM(\xi)\over d\xi}=\xi^{2}e^{-\psi}.
\label{em4}
\end{equation}
In addition, the metric functions determined by equations (\ref{eq10}) and (\ref{eq6}) can be expressed as
\begin{equation}
e^{-\lambda}=1-{2q\over 1+q}{M(\xi)\over\xi},\qquad \qquad \nu={2q\over 1+q}\psi(\xi)+{\it Cst.},
\label{em5}
\end{equation}
where the constant is determined by the matching with the outer Schwarzschild solution (\ref{eq13}) at $r=R$.

\begin{figure}[htbp]
\centerline{
\psfig{figure=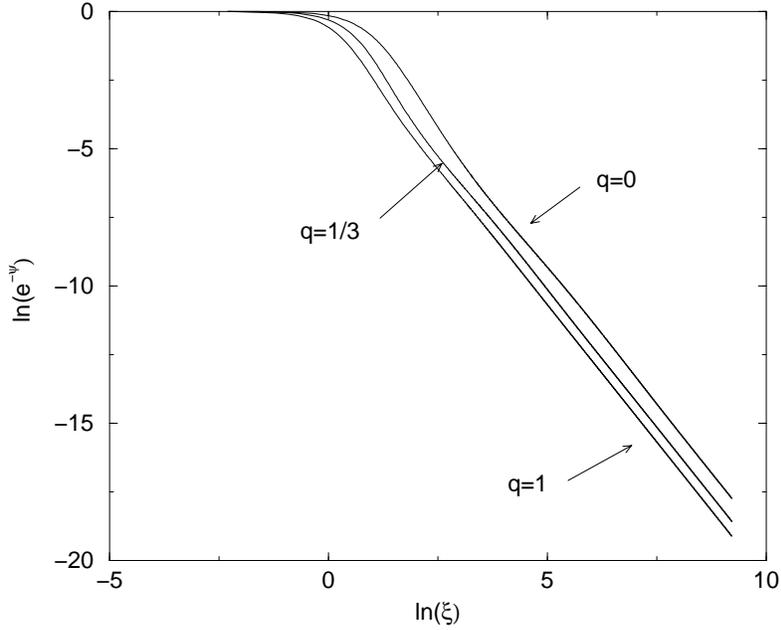,angle=0,height=8.5cm}}
\caption{Density profiles of the general relativistic Emden equation for different values of $q$. The profiles behaves like $\xi^{-2}$ at large distances. }
\label{density}
\end{figure}

The Newtonian limit corresponds to $q\rightarrow 0$. In that limit, Eqs. (\ref{em3})-(\ref{em4}) reduce to
\begin{equation}
{d\psi\over d\xi}={M(\xi)\over\xi^{2}},\qquad {\rm and} \qquad {dM(\xi)\over d\xi}=\xi^{2}e^{-\psi},
\label{em6}
\end{equation}
and they combine to give Emden's equation
\begin{equation}
{1\over\xi^{2}}{d\over d\xi}\biggl (\xi^{2}{d\psi\over d\xi}\biggr )=e^{-\psi}.
\label{em7}
\end{equation}
Therefore, Eqs.(\ref{em3})-(\ref{em4}) represent the general relativistic equivalent of the Emden equation. Like in the Newtonian case, they admit a singular solution of the form (Chandrasekhar 1972)
\begin{equation}
e^{-\psi_{s}}={Q\over \xi^{2}}, \qquad {\rm where}\qquad Q={2(1+q)\over (1+q)^{2}+4q}.
\label{em8}
\end{equation}
The metric associated with the singular isothermal sphere is given explicitly by
\begin{equation}
e^{\nu}=A\xi^{4q\over 1+q},\qquad e^{\lambda}=1+{4q\over (1+q)^{2}},
\label{em9}
\end{equation}
where $A$ is an unimportant constant. Considering now the regular solutions of  Eqs. (\ref{em3})-(\ref{em4}), we can always suppose that $\epsilon_{0}$ represents the energy density at the center of the configuration. Then, Eqs. (\ref{em3})-(\ref{em4}) must be solved with the boundary conditions
\begin{equation}
\psi(0)=\psi'(0)=0.
\label{em10}
\end{equation}
These solutions  must be computed numerically. However, it is possible to determine their asymptotic behaviors explicitly. For $\xi\rightarrow 0$, 
\begin{equation}
\psi=a\xi^{2}+b\xi^{4}+...,
\label{em11}
\end{equation}
with
\begin{equation}
a={1+3q\over 6},\qquad \qquad b=-{(15 q^{2}-2q+3)(1+3q)\over 360 (1+q)},
\label{em12}
\end{equation}
and for $\xi\rightarrow +\infty$ (Chandrasekhar 1972)
\begin{equation}
e^{-\psi}={Q\over\xi^{2}}\biggl\lbrace 1+{A\over \xi^{(1+3q)/ 2(1+q)}}\cos\biggl ({(7+42q-q^{2})^{1/2}\over 2(1+q)}\ln\xi+\delta\biggr )\biggr\rbrace.
\label{em13}
\end{equation}
The curve (\ref{em13}) intersects the singular solution (\ref{em8}) infinitely often at points that asymptotically increase geometrically in the ratio $1:{\rm exp}{{2\pi (1+q)\over (7+42q-q^{2})^{1/2}}}$ (see Fig. \ref{density}). Since the energy density $\epsilon$ falls off as $\xi^{-2}$ at large distances, the total mass-energy $M$ is {infinite}. In practice, this ``infinite mass problem'' does not arise because the isothermal equation of state (\ref{S6}) only holds in a finite region of space (e.g., the inner regions of a neutron star). For simplicity, we shall remedy this difficulty by assuming that our system is confined within a box of radius $R$, like in section \ref{sec_antonov}. This is clearly an idealization but it provides a well-posed model which captures the essential features of the system (see below) and which can be be studied in great detail without any further approximation.

If the system is enclosed within a box, the solutions of Eqs. (\ref{em3})(\ref{em4}) must be terminated at different radii $\alpha$ given by
\begin{equation}
\alpha=\biggl\lbrace {4\pi G\epsilon_{0}(1+q)\over c^{4}q}\biggr \rbrace^{1/2}R.
\label{em14}
\end{equation}
It should be noted that $\alpha$ is a measure of the central energy density $\epsilon_{0}$. Instead of $\alpha$, we might prefer to consider the density contrast 
\begin{equation}
{\cal R}\equiv {\epsilon_{0}\over\epsilon(R)}=e^{\psi(\alpha)}.
\label{em15}
\end{equation} 
The density contrast is a monotonic function of $\alpha$ varying from ${\cal R}=1$ (homogeneous system, $\alpha=0$) to ${\cal R}\rightarrow +\infty$ (singular sphere, $\alpha\rightarrow +\infty$). It is sometimes of interest to express the results in terms of the redshift (see, e.g., Weinberg 1972)
\begin{equation}
z={\Delta\lambda\over\lambda}=|g_{00}(r)|^{-1/2}-1=e^{-\nu(r)/2}-1.
\label{em16}
\end{equation} 
Using Eq. (\ref{em5}) with the boundary condition (\ref{eq13}) at $r=R$, the redshift of a spectral line emitted from an isothermal sphere is given by
\begin{equation}
z(\xi)=\biggl (1-{2GM\over Rc^{2}}\biggr )^{-1/2}{\rm exp}\biggl\lbrace {q\over 1+q}(\psi(\alpha)-\psi(\xi))\biggr\rbrace -1.
\label{em17}
\end{equation}

\subsection{The Milne variables}
\label{sec_milneg}

As in the Newtonian theory, it will be convenient in the following to introduce the Milne variables 
\begin{equation}
u={\xi e^{-\psi}\over \psi'},\qquad {\rm and} \qquad v=\xi\psi'.
\label{mg1}
\end{equation}
In terms of these variables, the system of equations (\ref{em3})(\ref{em4}) can be reduced to a single first order differential equation (Chandrasekhar 1972)
\begin{equation}
{u\over v}{dv\over du}={-1-{2q\over 1+q}v+(1+3q)u+{q(3+5q)\over 1+q}uv+{2q^{2}(1-q)\over (1+q)^{2}}uv^{2}\over 3-{1-q\over 1+q}v-(1+3q)u-{q(3+q)\over 1+q}uv-{4q^{2}\over (1+q)^{2}}uv^{2}}.
\label{mg2}
\end{equation}
For $\xi\rightarrow 0$, one has
\begin{equation}
u={1\over 2a}-\bigl ({b\over a^{2}}+{1\over 2}\bigr )\xi^{2}+...,\qquad \qquad v=2a\xi^{2}+...,
\label{mg3}
\end{equation}
and for $\xi\rightarrow +\infty$
\begin{equation}
u\rightarrow u_{s}={Q\over 2},\qquad\qquad v\rightarrow v_{s}=2.
\label{mg4}
\end{equation}

\begin{figure}[htbp]
\centerline{
\psfig{figure=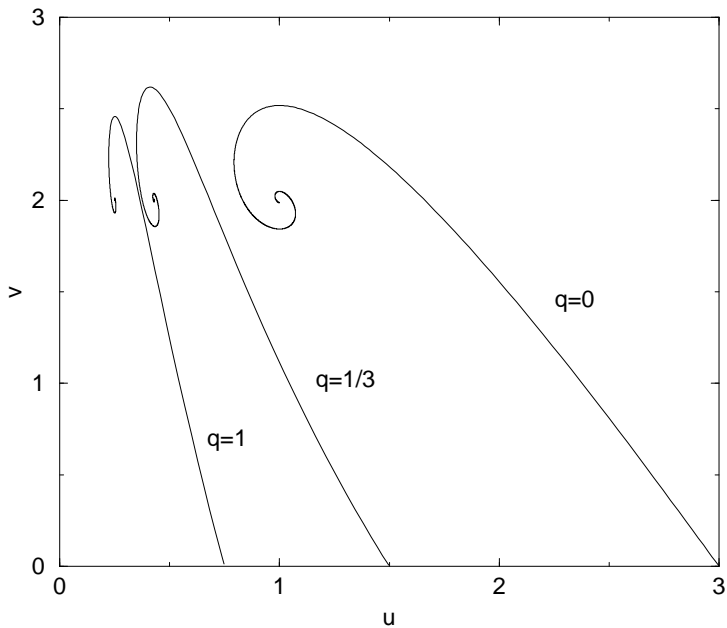,angle=0,height=8.5cm}}
\caption{The $(u,v)$ plane for isothermal gaseous spheres in general relativity and for different values of $q$. The value $q=0$ corresponds to the Newtonian limit. }
\label{uvq}
\end{figure}

The solution curve in the $(u,v)$ plane is parametrized by $\xi$. Starting from the point $(u,v)=({3\over 1+3q},0)$ for $\xi=0$ with a slope $(dv/du)_{0}=-{4 a^{3}\over a^{2}+2b}$, the solution curve spirals indefinitely around the point $(u_{s},v_{s})=({Q\over 2},2)$, corresponding to the singular sphere,  as $\xi\rightarrow +\infty$. All isothermal spheres must necessary lie on this curve (see Fig. \ref{uvq}). For bounded isothermal spheres, $\xi$ must be terminated at the box radius $\alpha$. Clearly, the spiral behavior of the $(u,v)$ curve can be ascribed to the oscillating behavior of the solution (\ref{em13}) as $\xi\rightarrow +\infty$. An explicit equation for the spiral (valid for $\xi\rightarrow +\infty$) can be obtained by substituting the asymptotic expansion (\ref{em13}) in the Milne variables (\ref{mg1}) like in the Newtonian case (see Chavanis 2001).

\subsection{Oscillatory behavior of the Mass-density profile}
\label{sec_osci}

According to Eq. (\ref{em2}), the relation between the total mass $M$ of the configuration and the central energy density $\epsilon_{0}$ (through the parameter $\alpha$) is given by
\begin{equation}
M={q\over 1+q}{M(\alpha)\over\alpha}{Rc^{2}\over G}.
\label{o1}
\end{equation}
Solving for $M(\xi)$ is Eq. (\ref{em3}), we get
\begin{equation}
{M(\alpha)\over \alpha}={\alpha\psi'(\alpha)-q\alpha^{2}e^{-\psi(\alpha)}\over 1+p\alpha \psi'(\alpha)},  \qquad \biggl (p={2q\over 1+q}\biggr ).
\label{o2}
\end{equation}
This relation can be expressed very simply in terms of the values of the Milne variables  $u$ and $v$ at the normalized box radius $\alpha$. Writing $u_{0}=u(\alpha)$ and $v_{0}(\alpha)$ and using Eq. (\ref{mg1}), we obtain 
\begin{equation}
\chi\equiv {2GM\over Rc^{2}}={pv_{0}(1-qu_{0})\over 1+pv_{0}}.
\label{o3}
\end{equation}

\begin{figure}[htbp]
\centerline{
\psfig{figure=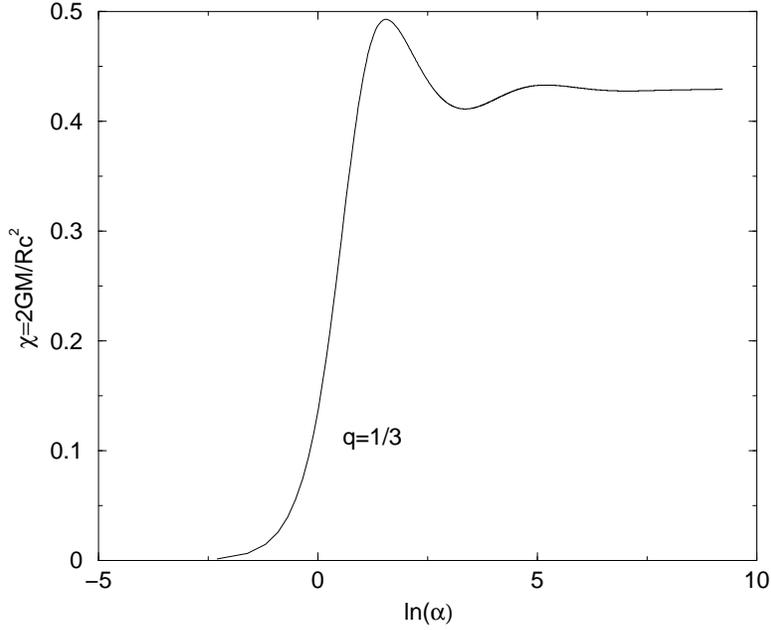,angle=0,height=8.5cm}}
\caption{Total mass-energy  $\chi$ versus central energy density $\alpha$ for $q=1/3$.}
\label{chialpha}
\end{figure}

The curve $\chi(\alpha)$ starts from $\chi=0$ for $\alpha=0$ and oscillates around its asymptotic value  $\chi_{s}=pQ= {4q\over (1+q)^{2}+4q}$ (corresponding to the singular sphere) as $\alpha\rightarrow +\infty$ (see Fig. 
\ref{chialpha}). This relation between the total mass and the central density is very similar to the corresponding one for neutron star models (Oppenheimer \& Volkoff 1939, Misner \& Zapolsky 1964) and dense stellar clusters at statistical equilibrium (Bisnovatyi {\it et al.} 1998). In particular, there exists mass peaks in the diagram. In fact, this oscillatory behavior should not cause surprise since the same phenomenon exists for a classical isothermal gas [see in particular Fig. 7 of Chavanis (2001) giving the mass-density relation for a fixed temperature].

\begin{figure}[htbp]
\centerline{
\psfig{figure=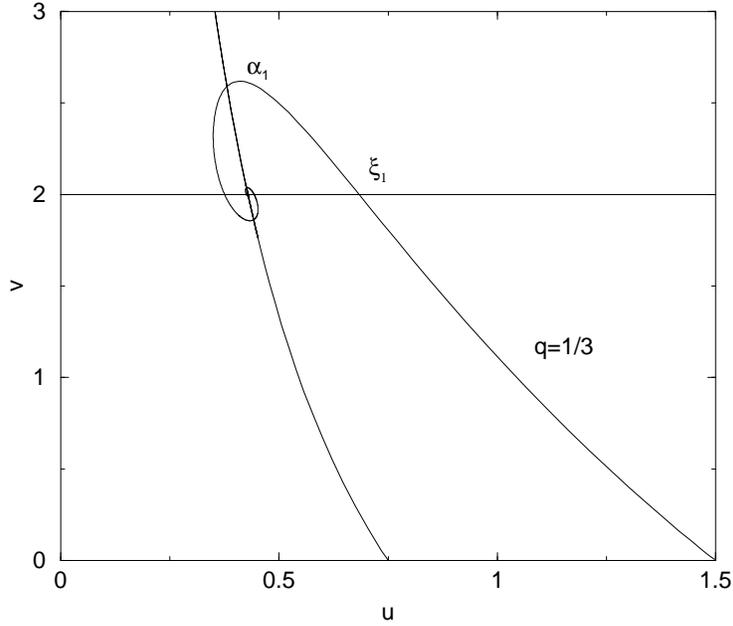,angle=0,height=8.5cm}}
\caption{Graphical construction to determine the critical density 
$\alpha_{1}$ for $q=1/3$. }
\label{uvchi}
\end{figure}

From Eqs. (\ref{o1})(\ref{o2}) and (\ref{em4}), one has
\begin{eqnarray}
{d\chi\over d\alpha}=p{M'(\alpha)\over\alpha}-p{M(\alpha)\over\alpha^{2}}=p \alpha e^{-\psi(\alpha)}-p {\psi'(\alpha)-q\alpha e^{-\psi(\alpha)}\over 1+ p\alpha\psi'(\alpha)}.
\label{o4}
\end{eqnarray}
In terms of the Milne variables, it can be rewritten
\begin{eqnarray}
{d\chi\over d\alpha}={p\over\alpha}\biggl (u_{0}v_{0}-{v_{0}(1-qu_{0})\over 1+pv_{0}}\biggr ).
\label{o5}
\end{eqnarray}
Therefore, the extrema of the curve $\chi(\alpha)$, determined by the condition $d\chi/d\alpha=0$, satisfy
\begin{equation}
p v_{0}={1\over u_{0}}-q-1.
\label{o6}
\end{equation}
This equation defines a hyperbole in the $(u,v)$ plane. For $u_{0}\rightarrow 0^{+}$, $v_{0}\sim {1\over p u_{0}}\rightarrow +\infty$ and for $u_{0}\rightarrow +\infty$, $v_{0}\rightarrow -{(q+1)^{2}\over 2q}$. The intersections between this curve and the spiral (see Fig. \ref{uvchi}) determines the values of $\alpha$ for which $\chi$ is an extremum. Since the curve (\ref{o6}) passes through the singular sphere $(u_{s},v_{s})$, there is an infinity of intersections with the spiral, resulting in an infinity of oscillations in Fig. \ref{chialpha}.

The first value $\alpha_{1}$ corresponds to the global maximum of $\chi$. Therefore, isothermal spheres in general relativity exist only provided that
\begin{equation}
\chi\le \chi_{c}\equiv {p v(\alpha_{1})(1-qu(\alpha_{1}))\over 1+pv(\alpha_{1})}. 
\label{o7}
\end{equation}
This implies in particular that, for a given radius, confined isothermal gas spheres exist only below a limiting mass
\begin{equation}
M<M_{c}\equiv \chi_{c}{Rc^{2}\over 2G}.
\label{o8}
\end{equation}
This result is of course related to the existence of a limiting mass for neutron stars as evidenced by Oppenheimer \& Volkoff (1939). In their study, the value of the radius $R$ is determined by a proper modelling of the envelope. However, our results indicate that the essential properties of neutron stars: oscillatory behavior of the mass-density profile, limiting mass and spiral behavior of the  mass-radius diagram (see section \ref{sec_mr}) are due primarily to their isothermal core and not to their envelope. Similar observations have been made by Yabushita (1974) who considered the case of an isothermal gaseous sphere surrounded by a medium exerting on it a constant pressure.

\begin{figure}[htbp]
\centerline{
\psfig{figure=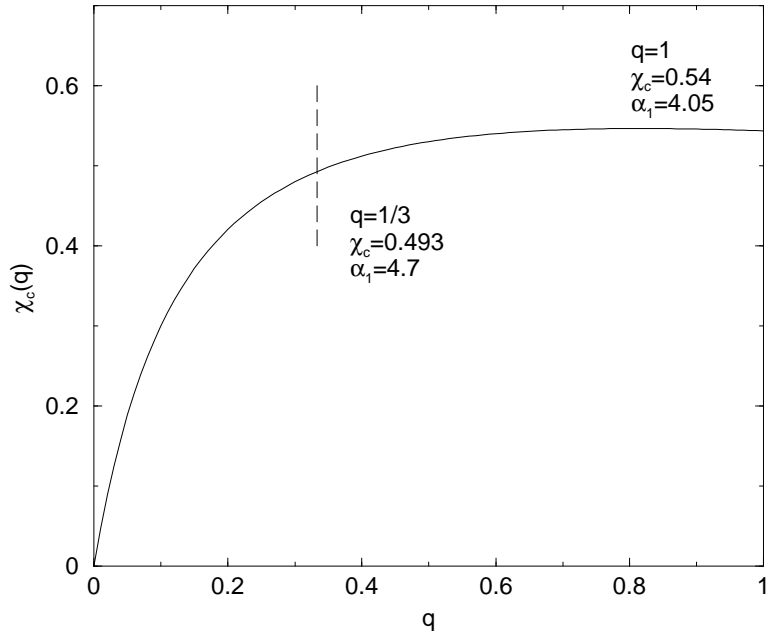,angle=0,height=8.5cm}}
\caption{Critical parameter $\chi_{c}$ as a function of $q$. The classical limit $q\rightarrow 0$ is discussed in section \ref{sec_length} }
\label{chicritq}
\end{figure}

\begin{figure}[htbp]
\centerline{
\psfig{figure=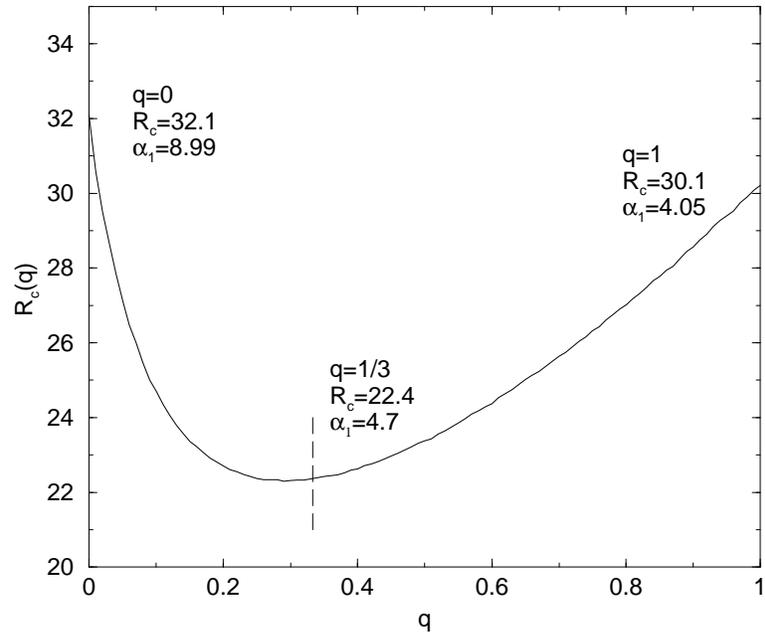,angle=0,height=8.5cm}}
\caption{Critical density contrast ${\cal R}_{c}$ as a function of $q$.}
\label{contrastq}
\end{figure}

\begin{figure}[htbp]
\centerline{
\psfig{figure=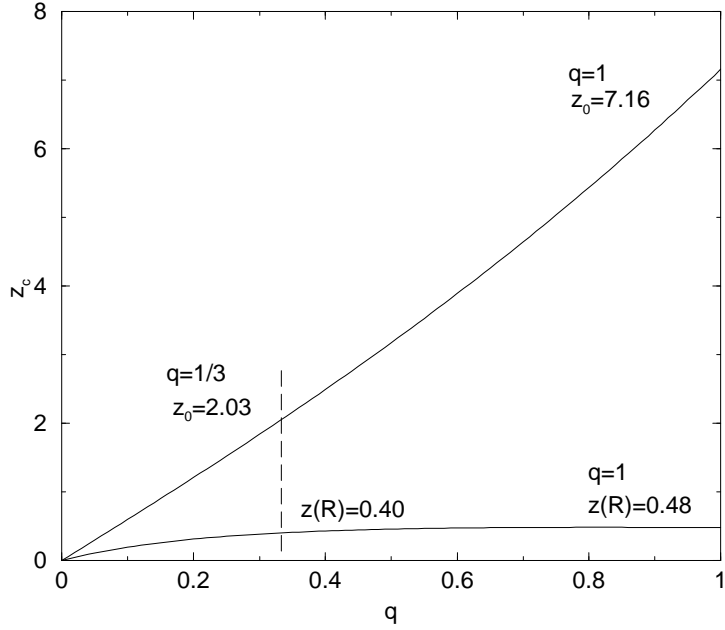,angle=0,height=8.5cm}}
\caption{Critical redshifts $z_{c}$ emitted from the center and from the boundary of an isothermal configuration as a  function of $q$.}
\label{redshift}
\end{figure}

The critical parameter $\chi_{c}$, the critical density contrast  ${\cal R}_{c}$ and the critical redshifts $z_{c}$  emitted from the center and from the boundary of the isothermal configuration are ploted as a function of $q$ on Figs. \ref{chicritq}-\ref{contrastq}-\ref{redshift}. The corresponding values of $\alpha_{1}$ are indicated on Fig. \ref{zero} (full line).  For the Newtonian case ($q=0$), we recover the classical values $\alpha_{1}=8.99$ and ${\cal R}_{c}=32.1$ obtained in the canonical ensemble. When $q$ is increased, the critical density $\alpha_{1}$ is lowered so that instability occurs sooner than in the Newtonian case. It should be noted, however, that the critical density contrast ${\cal R}_{c}$ is {\it not} a monotonous function of $q$. This is due to the deformation of the spiral in the $(u,v)$ plane when we vary the relativistic parameter $q$. We find that the value of $q$ for which ${\cal R}_{c}$ is minimum is close to $1/3$, the typical value corresponding to neutron stars. However, the variation of the critical density contrast with $q$ is not very important (${\cal R}_{c}\sim 20-30$ in the whole range of parameters)  so that the critical central redshift
\begin{equation}
z_{0}^{c}=(1-\chi_{c})^{-1/2}{\cal R}_{c}^{q\over 1+q}-1,
\label{newqt}
\end{equation}
obtained from Eqs. (\ref{em17}) (\ref{em15}) (\ref{o3}), increases  monotonically with $q$.   

\subsection{The mass-radius diagram}
\label{sec_mr}

It is well-known that, for high central densities, the mass-radius diagram for neutron stars presents a spiral behavior. We show in this section that we can reproduce this behavior within our simple ``box'' model. To that purpose, we shall consider configurations with different masses and radii but with the same density $\epsilon(R)$ at the boundary of the domain. According to Eqs. (\ref{em14}) and (\ref{em15}), we can express the radius $R$ as a function of the parameter $\alpha$ by the relation
\begin{equation}
{R\over R_{0}}=\alpha e^{-\psi(\alpha)/2},
\label{mr1}
\end{equation}
where we have introduced a typical radius
\begin{equation}
R_{0}=\biggl\lbrace{qc^{4}\over 4\pi G\epsilon(R)(1+q)}\biggr\rbrace^{1/2}.
\label{mr2}
\end{equation} 
On the other hand, using Eq. (\ref{o3}), the dimensionless mass appropriate to the present context is given by
\begin{equation}
{M\over M_{0}}=\chi(\alpha){R\over R_{0}}(\alpha),
\label{mr3}
\end{equation} 
where 
\begin{equation}
M_{0}={c^{2}R_{0}\over 2G}.
\label{mr4}
\end{equation}

\begin{figure}[htbp]
\centerline{
\psfig{figure=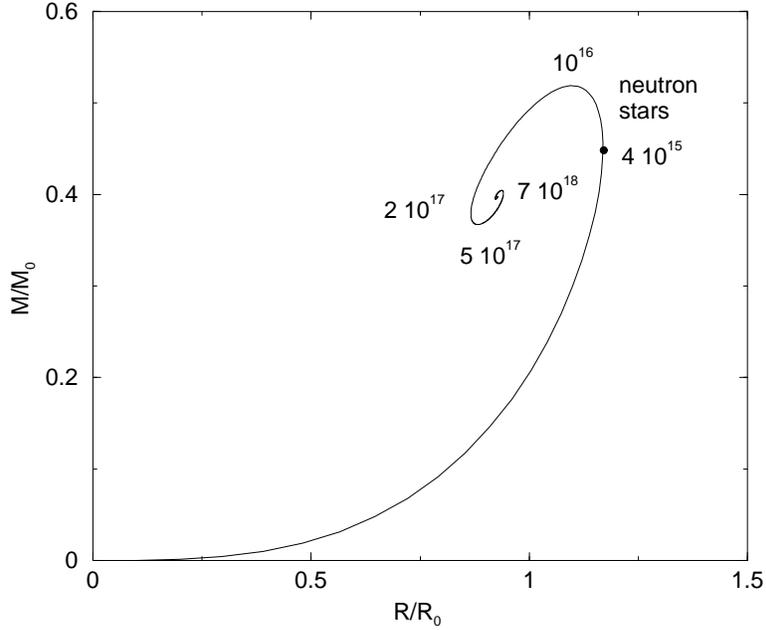,angle=0,height=8.5cm}}
\caption{The mass-radius diagram for an isothermal gas with a fixed energy density at the boundary ($\epsilon(R)=10^{15}gc^{2}/cm^{3}$). We have indicated the value of the central density $\epsilon_{0}$ at the turning points. }
\label{MassRadius}
\end{figure}

Eqs. (\ref{mr1})(\ref{mr3}) determine the mass-radius relation for an isothermal gas with a fixed energy density at its boundary. The $M-R$ curve, represented on Fig. \ref{MassRadius}, is parametrized by the central density $\alpha$.  For comparison with real objects (e.g., neutron stars) we note that the assumption of an isothermal core is valid only for sufficiently large central densities, typically from the point encircled. From that point, our model reproduces qualitatively the spiral behavior of the mass-radius diagram for neutron stars.  Using the asymptotic expansion (\ref{em13}) of the function $\psi(\xi)$, it is possible to determine an explicit equation for the spiral (valid for high densities) in the form
\begin{equation}
{R\over R_{0}}=Q^{1/2}\biggl\lbrace 1+{A\over 2\alpha^{(1+3q)/ 2(1+q)}}\cos\biggl ({(7+42q-q^{2})^{1/2}\over 2(1+q)}\ln\alpha+\delta\biggr ) \biggr\rbrace,
\label{mr5}
\end{equation} 
\begin{eqnarray}
{M\over M_{0}}=pQ^{3/2}\biggl\lbrace 1+{A\over 4(1+5q)\alpha^{(1+3q)/ 2(1+q)}}
\biggl\lbrack (1+q)(7+42q-q^{2})^{1/2}\qquad\qquad \nonumber\\
\qquad  \times\sin\biggl ({(7+42q-q^{2})^{1/2}\over 2(1+q)}\ln\alpha+\delta\biggr )
+(3+10q-q^{2})\cos \biggl ({(7+42q-q^{2})^{1/2}\over 2(1+q)}\ln\alpha+\delta\biggr ) \biggr\rbrack  \biggr\rbrace.\nonumber\\
\label{mr6}
\end{eqnarray} 
In particular, the terminal point of the spiral, corresponding to the singular solution (\ref{em8}), is given by
\begin{equation}
{R_{s}\over R_{0}}=Q^{1/2},\qquad {M_{s}\over M_{0}}=pQ^{3/2}.
\label{mr7}
\end{equation} 
The previous results are valid for any value of the parameter $q$.
We shall now specialize on the case of neutron stars
for which $q=1/3$. For estimating $\epsilon(R)$, we
shall adopt the typical value given by Misner \& Zapolsky (1964)
\begin{equation}
\epsilon(R)= 10^{15}\ gc^{2}/cm^{3},
\label{mr8}
\end{equation} 
which corresponds in an ideal neutron gas to a Fermi energy of $130$ MeV. 
This is the typical value of the energy density in the regime where the kinetic energy of the neutrons is of the same order as their rest mass. This regime corresponds to the transition between the isothermal core and the envelope and it is therefore relevant to adopt this value at $r=R$. For this density,   
\begin{equation}
R_{0}=5.2\ {\rm km},\qquad M_{0}=1.8\ M_{\odot}.
\label{mr9}
\end{equation} 
From the diagram of Fig. \ref{MassRadius}, we find that the maximum mass and the corresponding radius have the values
 \begin{equation}
R_{c}=1.1R_{0},\qquad M_{c}=0.5M_{0}.
\label{mr10}
\end{equation} 
The value of the maximum mass in our model $M_{c}\sim 0.9M_{\odot}$ is close to the more exact value $\sim 0.7M_{\odot}$ found by Oppenheimer \& Volkoff (1939) by introducing a more general equation of state valid also in the envelope. The value of the corresponding radius  $R_{c}=6\ km$ is slightly smaller than the value $\sim 9.6\ km$  usually reported (see, e.g., Weinberg 1972) and the difference in attributed in part to the presence of the envelope which surrounds the isothermal region of size $R_{c}$.  According to Eq. (\ref{em17}), the fractional redshift of a spectral line emitted from the edge of the isothermal region (for the configuration of maximum mass) is 
 \begin{equation}
z^{c}(R)=\biggl (1-{2GM_{c}\over R_{c}c^{2}}\biggr )^{-1/2}-1=0.35.
\label{mr11}
\end{equation} 
and the corresponding central redshift $z^{c}_{0}=1.4$. Finally, we find that the central density at the mass peak is $\epsilon_{0}^{c}=10^{16}gc^{2}/cm^{3}$ is good agreement with the reference model of Meltzer \& Thorne (1966) giving a comparable value. Other central densitites are indicated on Fig. \ref{MassRadius}. Our curve matches relatively well (qualitatively and semi-quantitatively) the diagram of Meltzer \& Thorne (1966) for central densities $\epsilon_{0}\gtrsim 4\ 10^{15}gc^{2}/cm^{3}$.

\section{Dynamical instability of isothermal gas spheres in general relativity}
\label{sec_dyn} 

\subsection{The condition of marginal stability}
\label{sec_con} 

The stability of relativistic stars against radial perturbations is
usually studied with the equation of pulsations derived by
Chandrasekhar (1964). Then, the stability of stars can be reduced to a
Sturm-Liouville problem, which is usually solved numerically (see,
e.g., Bardeen {\it et al.} 1966). The method is quite general but it
is not the most efficient in the present situation. Indeed, by
formulating the problem differently, using the form of the equation of
pulsations given by Yabushita (1973) and making use of the Milne
variables, it is possible to study the stability of bounded isothermal
spheres analytically, like in the Newtonian case.

Let us consider a small perturbation $\delta\epsilon$ around a configuration of isothermal equilibrium and let us introduce the mass perturbation $f(\xi)$ within the sphere of radius $\xi$ such that
\begin{equation}
{\delta\epsilon\over\epsilon_{0}}={1\over 4\pi\xi^{2}}{df\over d\xi}.
\label{con1}
\end{equation} 
Yabushita (1973,1974) has shown that the equation for radial pulsations satisfied by the function $f$ can be written 
\begin{equation}
{d^{2}f\over d\xi^{2}}+\biggl (-{2\over\xi}+{d\psi\over d\xi}+q\xi e^{\lambda-\psi}\biggr ){df\over d\xi}+e^{\lambda-\psi}\biggl \lbrack 1+{2q\xi\over 1+q}{d\psi\over d\xi}\biggr\rbrack f={\sigma^{2}c^{4}\over 4\pi G\epsilon_{0}(1+q)}e^{\lambda-\nu}f,
\label{con2}
\end{equation}
where $\sigma$ is the period of oscillations defined by $\delta\epsilon\sim e^{\sigma ct}$. Clearly, the condition of instability corresponds to $\sigma^{2}>0$. On the other hand, the velocity profile of the perturbation is given by 
\begin{equation}
\delta u^{1}=-{\sigma c\over q+1}{1\over 4\pi \epsilon}{f\over r^{2}}e^{-\nu/2}.
\label{P6}
\end{equation}
If the system is confined within a box, we must require that $\delta u^{1}=0$ at $r=R$. The function $f(\xi)$ must therefore satisfy the boundary conditions
\begin{equation}
f(0)=f(\alpha)=0.
\label{con3}
\end{equation}
These boundary conditions are equivalent to the conservation of mass-energy $M$. In the Newtonian limit $q\rightarrow 0$, the equation of pulsations (\ref{con2}) reduces to the one considered by Yabushita (1968) and Chavanis (2001). We shall be particularly interested by the point of marginal stability $(\sigma=0)$ in the series of equilibrium at which the system becomes unstable. We then have to solve the equation
\begin{equation}
{d^{2}F\over d\xi^{2}}+\biggl (-{2\over\xi}+{d\psi\over d\xi}+q\xi e^{\lambda-\psi}\biggr ){dF\over d\xi}+e^{\lambda-\psi}\biggl \lbrack 1+{2q\xi\over 1+q}{d\psi\over d\xi}\biggr\rbrack F=0,
\label{con4}
\end{equation}
with $F(0)=F(\alpha)=0$.  If we denote by ${\cal L}$ the differential operator that appears in Eq. (\ref{con4}), it is possible to show that (Yabushita 1974)
\begin{equation}
{\cal L}(M(\xi))={\cal L}(\xi^{3}e^{-\psi})={d\psi\over d\xi}{dM\over d\xi}+{2q\xi\over 1+q}e^{\lambda-\psi}M(\xi){d\psi\over d\xi}.
\label{con5}
\end{equation}
Therefore, the eigenfunction associated with $\sigma=0$ is
\begin{equation}
F(\xi)=c_{1}(\xi^{3}e^{-\psi}-M(\xi)).
\label{con6}
\end{equation}
In the classical limit, we recover the result
$\xi^{3}e^{-\psi}-\xi^{2}\psi'$ derived from the analysis of the
second order variations of the free energy (in the canonical ensemble)
or from the Navier-Stokes equations (Chavanis 2001). The condition
$F(0)=0$ is automatically satisfied. The condition $F(\alpha)=0$
determines the values of $\alpha$ at which a new mode of stability is
lost.  According to Eq. (\ref{con6}), we get
\begin{equation}
M(\alpha)=\alpha^{3}e^{-\psi(\alpha)}.
\label{con7}
\end{equation}
Using Eq. (\ref{o2}) and introducing the Milne variables (\ref{mg1}), it is straightforward to check that this relation is equivalent to 
\begin{equation}
pv_{0}={1\over u_{0}}-q-1,
\label{con8}
\end{equation}
which is precisely Eq. (\ref{o6}).  Therefore, a new eigenvalue $\sigma^{2}$ becomes positive at the points where $\chi$ is extremum. In particular, the first instability sets in precisely at the point of maximum mass. Isothermal configurations with a central density $\alpha>\alpha_{1}$, a density contrast ${\cal R}>{\cal R}_{c}$ or a central redshift $z_{0}>z_{0}^{c}$ are unstable with respect to radial perturbations. Similar results were obtained by Misner \& Zapolsky (1964) who numerically solved the Chandrasekhar equation for radial pulsations applied to their neutron star model. Their method was, however, limited to the fundamental mode and they could not demonstrate that secondary mass peaks correspond to the change of sign for the next eigenvalues (see their discussion). Our method gives an unambiguous definite answer to that question, at least in the framework of our model. It has to be noted that these results differ from those obtained by Yabushita (1974) who found that instability occurs {\it before} the first mass peak (except in the Newtonian case $q=0$). Now, the three models only differ in the boundary conditions: (i) Yabushita's gas spheres have envelopes which exert a constant pressure. (ii) In our model, the volume of the isothermal region is assumed to be fixed. (iii) Misner \& Zapolsky introduce a more realistic envelope in which the equation of state has a physically admissible form. Our results suggest that imposing a fixed volume (Antonov point of view) is a better approximation than imposing a fixed pressure (Bonnor point of view) since it provides the same condition of stability as the reference model of Misner \& Zapolsky (1964) and other works.

\subsection{Perturbation profiles and secondary instabilities}
\label{sec_pert} 
 
The values of $\alpha$ at which a new mode of instability occurs are determined  by Eq. (\ref{con8}). For large values of $\alpha$ (high order modes), we can introduce the asymptotic expansion (\ref{em13}) in the Milne variables. Then, after some algebra, we find that the values of $\alpha_{n}$  at which a new instability occurs are asymptotically given by
\begin{equation}
{(7+42q-q^{2})^{1/2}\over 2(1+q)}\ln\alpha_{n}+\delta=\arctan\biggl ({q^{2}+20q+3\over (1+q)(7+42q-q^{2})^{1/2}}\biggr )+n\pi.
\label{P1}
\end{equation}
In continuity with the Newtonian limit, they follow the geometric progression
\begin{equation}
\alpha_{n}\sim \biggl\lbrack e^{2\pi (1+q)\over (7+42q-q^{2})^{1/2}}\biggr\rbrack^{n}\qquad (n\rightarrow +\infty, {\rm integer}).
\label{P2}
\end{equation}
For $q=0,1/3,1$ the ratio in the geometric progression is respectively  $10.74$, $6.25$ and $6.13$. It is minimum for $q=7/11$ with a value $5.93$.

\begin{figure}[htbp]
\centerline{
\psfig{figure=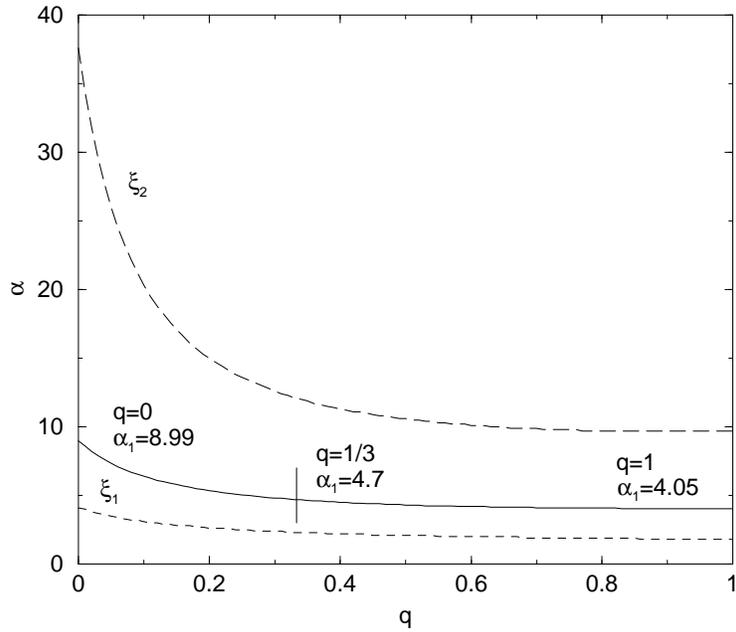,angle=0,height=8.5cm}}
\caption{Critical parameter $\alpha_{1}$ as a function of $q$ (full line) together with the first two zeros of the perturbation profile (\ref{P3}) satisfying $v(\xi_{i})=2$ (dash lines). Since $\xi_{2}>\alpha_{1}$, this solution must be rejected. }
\label{zero}
\end{figure}

\begin{figure}[htbp]
\centerline{
\psfig{figure=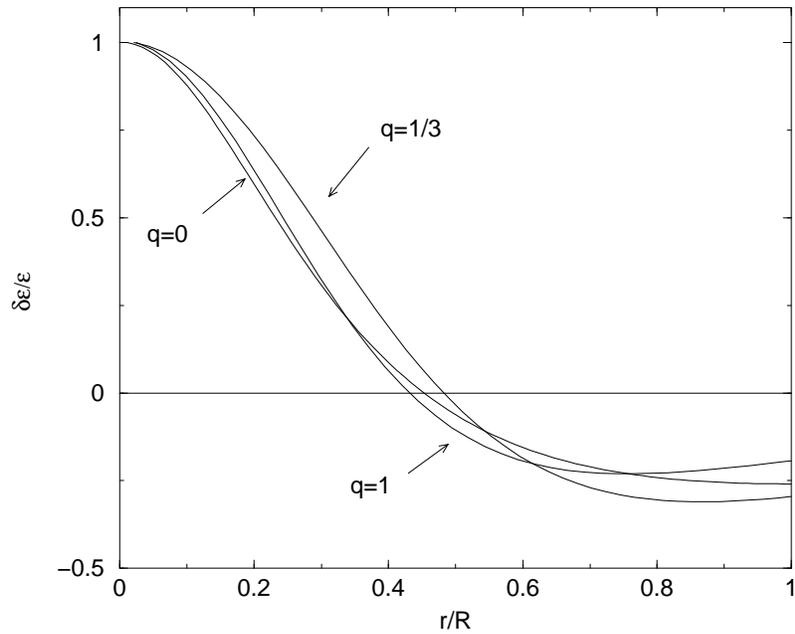,angle=0,height=8.5cm}}
\caption{Perturbation profiles at the point of marginal stability for different values of $q$. The perturbation profile does not present a core-halo structure. }
\label{pertuprof}
\end{figure}

The profile of the energy density perturbation associated with each instability is obtained by substituting Eq. (\ref{con6}) in Eq. (\ref{con1}). Simplifying the derivative with the aid of Eq. (\ref{em4}) and introducing the Milne variable $v$, we obtain
\begin{equation}
{\delta\epsilon\over\epsilon}={c_{1}\over 4\pi}(2-v),
\label{P3}
\end{equation}
The zeros of the profile $\delta\epsilon$ are determined by the intersections between the spiral in the $(u,v)$ plane and the curve $v=2$ (see Fig. \ref{uvchi}). The first two zeros are represented on Fig. 
\ref{zero} for different values of $q$. We see that $\xi_{2}$ is always larger than $\alpha_{1}$ so it does not satisfy the requirement  $0\le \xi\le \alpha$. Therefore, the first mode of instability has only one node at $\xi=\xi_{1}$. In other words, the perturbation profile that triggers the instability of an isothermal sphere in general relativity does not present a ``core-halo'' structure (see Fig. \ref{pertuprof}) in continuity with the Newtonian study in the canonical ensemble (Chavanis 2001).

The high order modes of instability have several nodes. Their positions (for large $\xi_{i}$) are obtained by expanding the Milne variable $v$ with the aid of Eq. (\ref{em13}) and substituting the resulting expression in the condition $v=2$. This yields
\begin{equation}
{(7+42q-q^{2})^{1/2}\over 2(1+q)}\ln\xi_{i}+\delta=-\arctan\biggl ({1+3q\over (7+42q-q^{2})^{1/2}}\biggr )+i\pi.
\label{P4}
\end{equation}
Therefore, the zeros of the perturbation profiles also follow a geometric progression
\begin{equation}
\xi_{i}\sim \biggl\lbrack e^{2\pi (1+q)\over (7+42q-q^{2})^{1/2}}\biggr\rbrack^{i}\qquad (i\rightarrow +\infty, {\rm integer}),
\label{P5}
\end{equation}
with the same ratio as the dimensionless radii $\alpha_{n}$. 

We can also determine the velocity profile (\ref{P6}) associated with each mode of instability. Introducing the velocity of sound $c_{s}=q^{1/2}c$ and expressing the result in terms of the dimensionless variables defined in section \ref{sec_emden}, we get
\begin{equation}
{\delta u^{1}\over c_{s}}=-{c\sigma' \over 4\pi(q+1)\xi^{2}}e^{\psi(\xi)\over 1+q}F(\xi),
\label{P7}
\end{equation}
where $F(\xi)$ is given by Eq. (\ref{con6}) at the critical points. In writing Eq. (\ref{P7}), it is implicitely understood that we are just at the {\it onset} of instability ($\sigma'=0^{+}$), so that Eq. (\ref{P7}) is applicable with $\sigma'>0$ (the velocity profile at the critical points, i.e. $\sigma'=0$, is simply $\delta u^{1}=0$).  In terms of the Milne variables, the velocity profile (\ref{P7}) can be written 
\begin{equation}
{\delta u^{1}\over c_{s}}=-{c\sigma' \over 4\pi(q+1)}c_{1}\psi'e^{\psi(\xi)\over 1+q}{u\over 1+pv}\biggl (pv-{1\over u}+q+1\biggr ).
\label{P8}
\end{equation}
For $q=0$, we recover the expression obtained in the Newtonian context from the Navier-Stokes equations (Chavanis 2001). Since $F(0)=F(\alpha)=0$, the velocity always vanishes at the center and at the boundary of the domain. The other zeros, for which $F(\xi_{i})=0$, correspond precisely to the values of $\alpha$ at which a new mode of instability occurs, i.e. $\xi_{i}=\alpha_{i}$. The $n^{th}$ mode of instability therefore has $n+1$ zeros $\xi=0$, $\xi=\alpha_{1}$, $\xi=\alpha_{2}$,..., $\xi=\alpha_{n}$. For high order modes of instability, the zeros follow asymptotically the geometric progression (\ref{P2}).

\subsection{The baryon number: another interpretation of the instability}
\label{sec_baryon} 

It has often been suggested on the basis of phenomenological arguments
that the baryon number $N$, or equivalently the rest mass $M_{0}=Nm$,
is also maximum at the onset of instability. We wish to check this
prediction explicitly in the case of an isothermal gas sphere.

The baryon number is obtained by multiplying the baryon number density $n$ by the {\it proper} volume element $e^{\lambda(r)/2}4\pi r^{2}dr$ and integrating the resulting expression  over the whole configuration. Using Eq. (\ref{eq10}), we get (see, e.g., Weinberg 1972) 
\begin{equation}
N=\int_{0}^{R}n\biggl\lbrack 1-{2GM(r)\over rc^{2}}\biggr\rbrack^{-1/2}4\pi r^{2}dr.
\label{N1}
\end{equation}
Now, according to the equations of state (\ref{S4})-(\ref{S6}), the baryon number density is related to the energy density by 
\begin{equation}
n^{\gamma}={q\over K}\epsilon.
\label{N2}
\end{equation}
Introducing the dimensionless variables defined in section \ref{sec_emden}, we find that the baryon number can be written
\begin{equation}
{N\over N_{*}}={1\over\alpha^{3q+1\over 1+q}}\int_{0}^{\alpha}e^{-{\psi(\xi)\over 1+q}}\biggl\lbrack 1-p{M(\xi)\over \xi}\biggr \rbrack^{-1/2}\xi^{2}d\xi,
\label{N3}
\end{equation}
with the normalization constant
\begin{equation}
N_{*}=4\pi R^{3}\biggl \lbrack {q^{2}c^{4}\over 4\pi GKR^{2}(1+q)}\biggr\rbrack^{1/\gamma}.
\label{N4}
\end{equation}

\begin{figure}[htbp]
\centerline{
\psfig{figure=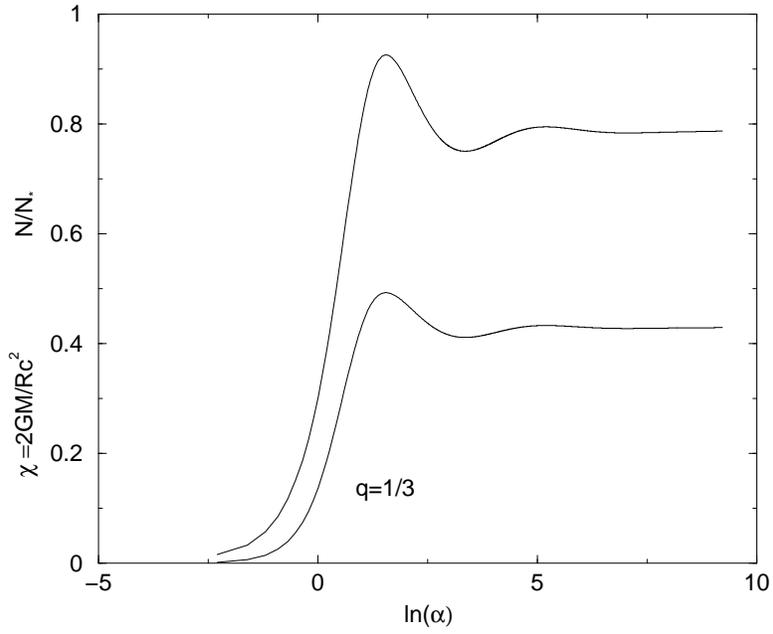,angle=0,height=8.5cm}}
\caption{Baryon number and total mass-energy as a function of the central density (through the parameter $\alpha$) for $q=1/3$. The peaks occur for the {same} values of the central density.}
\label{chiandN}
\end{figure}

\begin{figure}[htbp]
\centerline{
\psfig{figure=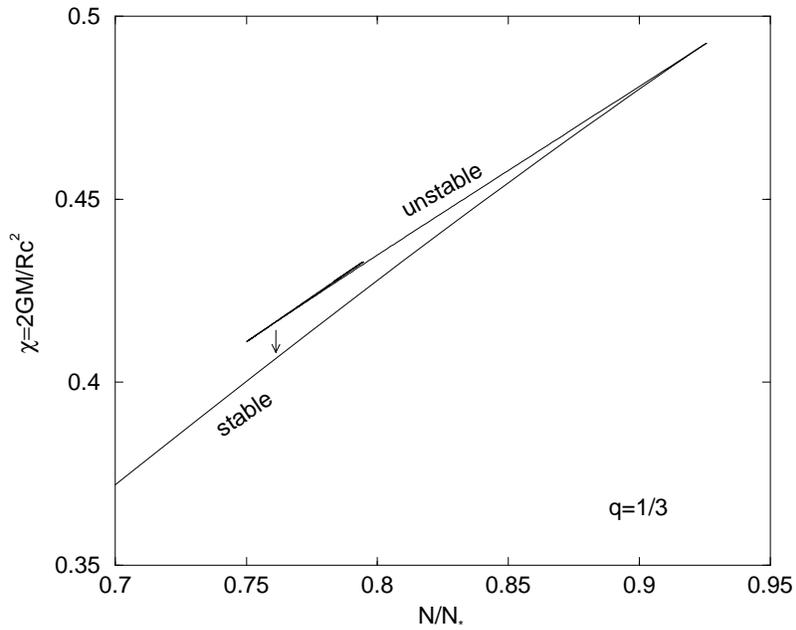,angle=0,height=8.5cm}}
\caption{Total mass-energy versus baryon number for isothermal gaseous spheres ($q=1/3$)}
\label{transition}
\end{figure}

The baryon number (\ref{N3}) is ploted versus the central density on
Fig. \ref{chiandN} for $q=1/3$. It exhibits damped oscillations, like
the total mass energy, and tends asymptotically to the value
${1+q\over 3q+1}Q^{1\over 1+q}(1-pQ)^{-1/2}$ as $\alpha\rightarrow
+\infty$ (singular sphere). It is observed {numerically} that the
peaks of mass-energy and of baryon number occur for the {\it same}
values of the central density (we have checked this property for
various values of $q$). This is also clear from Fig. \ref{transition}
where the mass-energy $M$ is ploted as a function of the baryon number
$N$. From this figure, we can understand physically why the system
becomes unstable after the first mass peak. Indeed, for a given rest
mass $Nmc^{2}$ there exist several solutions with different energy
$Mc^{2}$. Only the solution with the lowest energy (``fundamental''
state) is stable with respect to radial perturbations (see Weinberg
1972). We expect therefore some phase transitions to occur for a state
of high energy to a state of lower energy (see the arrow on
Fig. \ref{transition}). Therefore, only the lower branch is stable and
it corresponds to the configurations prior to the first mass peak (we
come to the same conclusion by minimizing the binding energy
$E=(M-M_{0})c^{2}$ at fixed mass $M$). These results are consistent
with those found by Tooper (1964) for polytropic gas spheres.

\subsection{Characteristic length scales}
\label{sec_length}

We shall now reformulate the previous results in a different form by introducing appropriate length scales. This will provide a direct comparison with our discussion in the Newtonian context (Chavanis 2001). In general relativity, a gaseous sphere of mass $M$ is unstable  if its radius is {\it smaller} than a multiple of the Schwarzschild radius (Chandrasekhar 1964). For a confined isothermal gas, this condition is explicitly given by Eq. (\ref{o8}) which can be rewritten
\begin{equation}
R\le {1\over \chi_{c}(q)}R_{S},\qquad R_{S}={2GM\over c^{2}}.
\label{L1}
\end{equation}
On the other hand, in Newtonian gravity, the condition of gravitational collapse is that the size of the domain be {\it larger} than the Jeans length $L_{J}$. For example, in the case of an isothermal gas confined within a box, the condition of instability reads (de Vega \& Sanchez 2001, Chavanis 2001)
\begin{equation}
R\ge \sqrt{2.52...\over 3}L_{J},\qquad L_{J}=\biggl ({3R^{3}\over \beta GMm}\biggr )^{1/2}.
\label{L2}
\end{equation}
It should be emphasized that these inequalities characterize the {\it
absence} of hydrostatic equilibrium for an isothermal sphere. This is
more stringent than the instability of an equilibrium configuration
(see below). Although the sense of these inequalities is opposite,
there is no contradiction since the size of the domain enters
explicitly in the definition of the Jeans length (through the average
density). For $q\rightarrow 0$, the two criteria become equivalent as
we now show. In the classical limit, the rest mass dominates the
kinetic energy so that $\epsilon=\rho c^{2}$ where $\rho=n m$ is the
classical mass density. Therefore, the equation of state (\ref{S6})
becomes $p=q\rho c^{2}$ and it must be compared with the equation of
state of a classical isothermal gas $p={\rho\over m}kT$. This yields
\begin{equation}
q={kT\over mc^{2}}={1\over\beta mc^{2}}\qquad (\beta mc^{2}\rightarrow +\infty).
\label{L3}
\end{equation}
Now, in the classical limit $c\rightarrow +\infty$, the critical parameter $1/\chi_{c}$, given by Eq. (\ref{o7}), diverges like
$1/\chi_{c}=1/2qv(\alpha_{1})\rightarrow +\infty$ while the Schwarzschild radius $R_{S}$ goes to zero in such a way that the product $(1/\chi_{c})R_{S}$ remains finite. Therefore, the inequality (\ref{L1}) is {\it reversed}
\begin{equation}
R\le {\beta G M m\over v(\alpha_{1})}={3R^{3}\over v(\alpha_{1})L_{J}^{2}}  \Longrightarrow R\ge \sqrt{v(\alpha_{1})\over 3}L_{J},
\label{L4}
\end{equation}
and we recover the result (\ref{L2}) with $v(\alpha_{1})=2.52$. It can be noted that the same criterion can be written
\begin{equation}
M\ge M_{c}\equiv 2.52 {RkT\over Gm}.
\label{L5}
\end{equation}  
Therefore, a classical gaseous sphere of radius $R$ and temperature $T$ can exist only below a limiting mass $M_{c}$ given by formula (\ref{L5}). Obviously, the physical origin of this limiting mass is the same as for the limiting mass of neutron stars but with a completely different interpretation of the parameters. 

It is also relevant to introduce another length scale  
\begin{equation}
L_{c}=\biggl ({9qc^{4}\over 4\pi G\epsilon_{0}(1+q)}\biggr )^{1/2}
\label{L6}
\end{equation} 
which gives a good estimate of the ``core radius'' of a relativistic isothermal gas sphere. In the classical limit, $L_{c}$ reduces to the King's length (Binney \& Tremaine 1987). In terms of the core radius, the parameter $\alpha$ can be written
\begin{equation}
\alpha=3{R\over L_{c}}
\label{L7}
\end{equation}  
The condition of instability $\alpha>\alpha_{1}$ is therefore equivalent to
\begin{equation}
R>{\alpha_{1}\over 3}L_{c}.
\label{L8}
\end{equation} 
Above this threshold, it is still possible to construct isothermal gas spheres in hydrostatic equilibrium but they are unstable. Secondary instabilities occur for radii that follow a geometric progression 
 \begin{equation}
R_{n}\sim\biggl\lbrack e^{2\pi (1+q)\over (7+42q-q^{2})^{1/2}}\biggr\rbrack^{n} L_{c}\qquad (n\rightarrow +\infty, {\rm integer}).
\label{L9}
\end{equation}
In the Newtonian context, this hierarchy of scales was interpreted by Semelin {\it et al.} (1999) and Chavanis (2001) in relation with the fragmentation of the interstellar medium and the possible origin of its fractal structure. 

The instability criteria (\ref{L1}) and (\ref{L8}), expressed in
terms of distinct length scales, can also be expressed in terms of
the redshift, with a similar distinction between the central redshift
and the surface redshift. According to Eq. (\ref{mr11}), the surface
redshift $z(R)$ is an increasing function of the mass energy $\chi$
while the central redshift $z_{0}$, deduced from Eq. (\ref{em17}), is
an increasing function of the central density $\alpha$. Therefore, the
surface redshift must necessarily satisfy $z(R)<z(R)^{c}$ since no
hydrostatic equilibrium of isothermal spheres can exist above this
threshold while the central redshift must satisfy
$z_{0}<z_{0}^{c}$ because configurations with $z_{0}>z_{0}^{c}$ exist
mathematically but are unstable.

\section{Conclusion}
\label{sec_conclusion}

In this paper, we have studied the structure and the stability of
isothermal gas spheres in the framework of general relativity. We have
found that the relativistic isothermal spheres (like neutron cores)
exhibit the same kind of behaviors as Newtonian isothermal spheres but
with a different interpretation of the parameters.  In this analogy,
the critical energy and the critical temperature found by Antonov
(1962) and Lynden-Bell \& Wood (1968) for an isothermal gas are the
classical equivalent of the limiting mass for neutron stars discovered
by Oppenheimer \& Volkoff (1939) [there also exists a limiting mass
for classical isothermal gas spheres, see Eq. (\ref{L5})]. Similarly,
the spiral behavior of the temperature-energy curve for a classical
gas has the same origin as the mass-radius diagram in neutron
stars. This spiral behavior, as well as the damped oscillations of the
mass-density profile, are general features of isothermal
configurations.

Although our model can reproduce qualitatively all the features of
neutron stars, it is however limited in its applications since the
isothermal core is surrounded by an artificial ``box'' instead of a
more physical envelope like in the studies of Oppenheimer \& Volkoff
(1939), Misner
\& Zapolsky (1964) and Meltzer \& Thorne (1966). However, this
simplification allows us to study the stability problem analytically
(or with graphical constructions) without any further
approximation. This is a useful complement to the more elaborate
works of the previous authors who had to solve the pulsation equation
numerically or with approximations. This is also complementary to the
study of Yabushita (1974) who considered isothermal gas spheres
surrounded by an envelope exerting a constant pressure. The boundary
conditions are of importance since Yabushita came to the conclusion
that the onset of instability occurs before the first mass peak. In
contrast, our model gives a stability criterion consistent with that
of Misner \& Zapolsky (1964) and other works.  Furthermore, if we
assume that the energy density at the edge of the isothermal core has
a value $\sim 10^{15}gc^{2}/cm^{3}$ (a typical prediction of nuclear
models), we can reproduce almost quantitatively the results usually
reported for neutron stars.

It should be stressed that the analogy between neutron stars and
isothermal spheres is a pure effect of general relativity. When
gravity is treated in the Newtonian framework, the classical condition
of hydrostatic equilibrium requires a relationship between the
pressure and the mass density $\rho=mn$ which, for dense matter
(degenerate and relativistic), is a power law with index $\gamma=4/3$
(see Chandrasekhar 1942). However, when gravity is treated in the
framework of general relativity, the Oppenheimer-Volkoff equations
require a relationship between the pressure and the mass-energy
density $\epsilon$ which, for dense matter, is linear. Therefore, the
core of neutron stars (treated with general relativity) is
``isothermal'' while the core of white dwarf stars approaching the
limiting mass (treated in a Newtonian framework)
is ``polytropic'', although the same equation of state is used (that
for a completely degenerate and ultra-relativistic ideal Fermi
gas). This is the intrinsic reason why the Mass-Radius relation for
neutron stars exhibits a spiral behavior (Meltzner \& Thorne 1966)
while the Mass-Radius relation for white dwarf stars is monotonous
(Chandrasekhar 1942). These remarks may increase the interest of studying
isothermal gas spheres both in Newtonian mechanics and in general
relativity.

\subsection{Acknowledgments}
\label{sec_ack}

This work was initiated during my stay at the Institute for Theoretical Physics, Santa Barbara, during the program on Hydrodynamical and Astrophysical Turbulence (February-June 2000). This research was supported in part by the National Science Foundation under Grant No. PHY94-07194.

\end{document}